%% file: 0_main.tex
\documentclass[10pt, journal, letterpaper]{IEEEtran}
\usepackage{amsmath,amsfonts}
\usepackage{array}
\usepackage[caption=false]{subfig}
\usepackage{textcomp}
\usepackage{stfloats}
\usepackage{url}
\usepackage{verbatim}
\usepackage{graphicx}

\usepackage[style=ieee, doi=false, url=false, isbn=false, giveninits=true, backend=biber, sorting=none, sortcites=true, date=year]{biblatex}
\DeclareSourcemap{
  \maps[datatype=bibtex]{
    \map{
      \step[fieldset=address, null]
      \step[fieldset=editor, null]
      \step[fieldset=location, null]
      \step[fieldset=primaryClass, null]
      \step[fieldset=publisher, null]
      \step[fieldset=series, null]
    }
  }
}
\usepackage{amsthm}
\usepackage{mathtools}
\usepackage{xcolor}
\definecolor{linkcolor}{rgb}{0.0, 0.0, 0.0}  
\definecolor{citecolor}{rgb}{0.0, 0.6, 0.3}  
\definecolor{urlcolor}{rgb}{0.6, 0.0, 0.3}   
\usepackage[hidelinks]{hyperref}

\usepackage{url}

\usepackage{booktabs}
\usepackage{multirow}
\usepackage{makecell}

\usepackage{algorithmicx}%
\usepackage{algorithm}
\usepackage{algpseudocode}%

\usepackage{bm}

\usepackage[nameinlink, sort, capitalize]{cleveref}
\Crefname{section}{Section}{Sections}
\Crefname{table}{Table}{Tables}

\usepackage[acronym]{glossaries}

\usepackage{tikz}
\usetikzlibrary{arrows.meta, positioning}

\usepackage{multicol}
\usepackage{setspace} 
\usepackage{mdframed} 
\newmdenv[
  linecolor=black,
  linewidth=1pt,
  topline=false,
  bottomline=false,
  rightline=false,
  skipabove=10pt,
  skipbelow=10pt,
]{customquote} 

\usepackage{wrapfig}
\usepackage{orcidlink}

\definecolor{Honeydew}{RGB}{240, 255, 240}
\definecolor{FT}{RGB}{255, 241, 229}
\newtheorem{problem}{Problem}
\newtheorem{corollary}{Corollary}

\newacronym{aep}{AEP}{Asymptotic Equipartition Property}
\newacronym{ai}{AI}{Artificial Intelligence}
\newacronym{aoi}{AoI}{Age of Information}
\newacronym{ar}{AR}{Augmented Reality}
\newacronym{awgn}{AWGN}{Additive White Gaussian Noise}
\newacronym{ber}{BER}{Bit-Error Rate}
\newacronym{bev}{BEV}{Bird's-eye View}
\newacronym{bpp}{BPP}{bits per pixel}
\newacronym{carla}{CARLA}{Car Learning to Act}
\newacronym{cav}{CAV}{Connected Autonomous Vehicle}
\newacronym{cps}{CPS}{Cyber-Physical Systems}
\newacronym{csi}{CSI}{Channel State Information}
\newacronym{cvae}{CVAE}{Conditional Variational Autoencoder}
\newacronym{dnn}{DNN}{Deep Neural Network}
\newacronym{dl}{DL}{Deep Learning}
\newacronym{dpgm}{DPGM}{Directed Probabilistic Graphical Model}
\newacronym{dtcp}{DTCP}{Delay-aware Trajectory-guided Control Prediction}
\newacronym{e2e}{E2E}{End-to-End}
\newacronym{fid}{FID}{Fréchet Inception Distance}
\newacronym{gan}{GAN}{Generative Adversarial Network}
\newacronym{gru}{GRU}{Gated Recurrent Unit}
\newacronym{ib}{IB}{Information Bottleneck}
\newacronym{jscc}{JSCC}{Joint Source-Channel Coding}
\newacronym{kl}{KL}{Kullback-Leibler}
\newacronym{ldpc}{LDPC}{Low-Density Parity-Check}
\newacronym{mdp}{MDP}{Markov Decision Process}
\newacronym{mse}{MSE}{Mean Square Error}
\newacronym{msssim}{MS-SSIM}{Multi-Scale Structural Similarity}
\newacronym{ofdm}{OFDM}{Orthogonal Frequency-Division Multiplexing}
\newacronym{psnr}{PSNR}{Peak Signal-to-Noise Ratio}
\newacronym{qam}{QAM}{Quadrature Amplitude Modulation}
\newacronym{rf}{RF}{Radio Frequency}
\newacronym{rma2b}{R$\text{(MA)}^2$B}{Restless Multi-Armed Multi-Action Bandit}
\newacronym{snr}{SNR}{Signal-to-Noise Ratio}
\newacronym{ssim}{SSIM}{Structural Similarity}
\newacronym{tgcp}{TGCP}{Trajectory-Guided Control Prediction}
\newacronym{tscc}{TSCC}{Task-oriented Source-Channel Coding}
\newacronym{urllc}{URLLC}{Ultra-Reliable Low-Latency Communication}
\newacronym{vae}{VAE}{Variational Autoencoder}
\newacronym{vib}{VIB}{Variational Information Bottleneck}
\newacronym{vr}{VR}{Virtual Reality}
\newacronym{v2x}{V2X}{Vehicle-to-Everything}
\newcommand*{\dif}{\mathop{}\!\mathrm{d}}

\glsdisablehyper

\begin{document}

\title{Task-Oriented Co-Design of Communication, Computing, and Control for Edge-Enabled Industrial Cyber-Physical Systems}

\author{
Yufeng~Diao,~\IEEEmembership{Graduate~Student~Member,~IEEE},~%
Yichi~Zhang,~\IEEEmembership{Graduate~Student~Member,~IEEE},~%
\\%
Daniele~De~Martini,~\IEEEmembership{Member,~IEEE},
Philip~Guodong~Zhao,~\IEEEmembership{Senior~Member,~IEEE},
and~Emma~Liying~Li,~\IEEEmembership{Member,~IEEE}%

\thanks{Yufeng Diao is with the School of Computing Science, University of Glasgow, U.K., and is currently a visiting Ph.D. student with the Department of Computer Science, University of Manchester, U.K. (e-mail: y.diao.1@research.gla.ac.uk).}%
\thanks{Yichi Zhang is with the Department of Computer Science, University of Manchester, U.K. Part of this work was done when he was with the James Watt School of Engineering, University of Glasgow, U.K. (e-mail: yichi.zhang@postgrad.manchester.ac.uk).}%
\thanks{Daniele De Martini is with the Oxford Robotics Institute, Department of Engineering Science, University of Oxford, U.K. (e-mail: daniele@robots.ox.ac.uk).}%
\thanks{Philip Guodong Zhao is with the Department of Computer Science, University of Manchester, U.K. (e-mail: philip.zhao@manchester.ac.uk).}%
\thanks{Emma Liying Li is with the School of Computing Science, University of Glasgow, U.K. (e-mail: liying.li@glasgow.ac.uk).}
}

\markboth{This paper has been accepted for publication in IEEE Journal on Selected Areas in Communications (JSAC).}%
{}


\maketitle

\begin{abstract}
\input{1_0_abstract}
\end{abstract}
\glsresetall

\input{1_1_introduction}

\input{1_2_related_work}
\input{2_1_IB}
\input{2_2_DTCP}

\input{3_experiment}
\input{4_conclusion}
\input{appendix}

\printbibliography

\end{document}

%% file: 1_0_abstract.tex
This paper proposes a task-oriented co-design framework that integrates communication, computing, and control to address the key challenges of bandwidth limitations, noise interference, and latency in mission-critical industrial Cyber-Physical Systems (CPS). To improve communication efficiency and robustness, we design a task-oriented Joint Source-Channel Coding (JSCC) using Information Bottleneck (IB) to enhance data transmission efficiency by prioritizing task-specific information. To mitigate the perceived End-to-End (E2E) delays, we develop a Delay-Aware Trajectory-Guided Control Prediction (DTCP) strategy that integrates trajectory planning with control prediction, predicting commands based on E2E delay. Moreover, the DTCP is co-designed with task-oriented JSCC, focusing on transmitting task-specific information for timely and reliable autonomous driving. Experimental results in the CARLA simulator demonstrate that, under an E2E delay of 1 second (20 time slots), the proposed framework achieves a driving score of 48.12, which is 31.59 points higher than using Better Portable Graphics (BPG) while reducing bandwidth usage by 99.19\%.

%% file: 1_1_introduction.tex
\section{Introduction}
\label{sec_intro}
In industrial \gls{cps}, ensuring \glspl{urllc} is crucial for achieving reliable real-time performance \cite{She_2021_ATo}. Applications such as automated transportation, material handling, and inspection increasingly rely on autonomous vehicles and robots within factories, warehouses, and hazardous environments. Autonomous driving plays an important role in these systems, enabling the automation of essential tasks, optimizing workflow efficiency, and improving safety \cite{Chen_2024_SMW}. 

To meet the stringent requirements of \gls{urllc} (e.g., the \gls{e2e} delay should be less than 1 ms and the packet loss probability should be less than \(10^{-5}\) \cite{3GPP}), edge inference has emerged as a promising solution \cite{She_2019_CLD}. By minimizing the physical distance between data generation and processing, edge inference significantly reduces latency, which is vital for autonomous systems that must respond timely to dynamic environmental changes, such as navigating unpredictable factory layouts or reacting to sudden obstacles. The primary motivation for using edge computing (also called off-board computing) lies in its ability to enhance system flexibility, particularly given the differing innovation cycles of the automotive and semiconductor industries. While a vehicle's lifespan typically ranges from 10 to 20 years, advancements in computing capabilities can be significant within this period. By leveraging edge computing within shared telecommunication infrastructure, the vehicle can enjoy much better flexibility in upgrading the computing power and software throughout their entire lifetime.
However, the transmission of massive amounts of sensor and video data presents a challenge to the edge’s ability to handle real-time processing while maintaining the reliability and low-latency communications. In that case, edge inference, often powered by \glspl{dnn}, is still affected by nontrivial communication latency and bandwidth constraints, particularly under the demands of \gls{urllc} in industrial \gls{cps} \cite{Khan_2022_Uae}.

Recent developments in deep learning have introduced \gls{jscc} as a promising solution to the limited communication bandwidth and significant noise interference \cite{Bourtsoulatze_2019_DJS}. Unlike traditional separate coding designs, \gls{jscc} integrates source and channel coding, improving data transmission efficiency. Despite these advantages, conventional \gls{jscc} approaches typically focus on accurate signal reconstruction at the receiver, potentially wasting communication resources on task-agnostic information that does not directly contribute to the decision-making process \cite{Kurka_2020_DfD}. This inefficiency has attracted significant research interest in task-oriented communication, a technology designed to prioritize the transmission of task-specific information, thus reducing data rates and improving efficiency, especially for AI-driven applications \cite{Diao_2024_TOS}. 

Furthermore, data sent to the edge server become outdated due to uplink delays, including processing, transmission, propagation, and queueing delay, negatively impacting the timeliness of edge inference results. This issue is further exacerbated by downlink, computing, and control delays. The \gls{e2e} delay (round-trip delay) degrades system performance and makes it difficult to meet the \gls{urllc} requirements in industrial \gls{cps}. Prediction-based methods can mitigate perceived \gls{e2e} delay \cite{Hou_2020_PaC}, but longer prediction horizons increase the risk of inaccuracy, creating a trade-off between minimizing delay and ensuring reliability.

Given the limitations of traditional approaches that design communication, computing, and control components separately, an integrated task-oriented co-design framework becomes essential \cite{Meng_2023_SCa, Meng_2024_TOC, Shao_2022_LTO}. In this work, our objective is to address three fundamental questions for edge-enabled mission-critical industrial \gls{cps}: 1) How can data transmission be optimized for bandwidth-constrained and latency-sensitive applications to ensure that task-specific information is prioritized? 2) How can predictive models be utilized to ensure that edge inference systems make decisions that reduce perceived \gls{e2e} delay? 3) How can communication, computing, and control be jointly designed and optimized to meet the demands of \gls{urllc} in mission-critical applications? The key contributions of this paper are summarized as follows:
\begin{itemize}
    \item We develop a comprehensive task-oriented co-design framework that jointly optimizes communication, computing, and control. This framework seamlessly integrates task-oriented \gls{jscc} with a delay-aware autonomous driving agent, addressing the critical challenges of bandwidth constraints, noise interference, and \gls{e2e} delay to maximize performance for edge-enabled autonomous driving.

    \item We formulate the problem of task-oriented communication using the \gls{ib} approach and employ a variational approximation to derive a tractable upper bound, resulting in the \gls{vib} method. Additionally, we extend the standard \gls{vib} framework to incorporate conditional information, such as vehicle and channel state information, ensuring better alignment with mission-critical applications. Furthermore, we handle the KL-divergence term using a concise approach inspired by \cite{Kingma_2013_Aev}. Our formulation improves communication efficiency in dynamic and noisy environments, which is essential for the reliable operation of industrial \gls{cps}.

    \item We establish the \gls{dtcp} strategy for autonomous driving, which uniquely combines two dominant autonomous driving paradigms: trajectory planning and control prediction. The \gls{dtcp} processes \gls{jscc} symbols, state information, and channel state to predict optimal driving actions that reduce perceived \gls{e2e} delay. In addition, \gls{dtcp} is specially co-designed with the task-oriented \gls{jscc} and is jointly trained for machine-to-machine communication.
\end{itemize}

The rest of this paper is organized as follows.
The related works are presented in \cref{sec_related_works}. In \cref{sec_system}, we introduce the system model and formulate the variational problem. \Cref{sec_DTCP} presents the details of \gls{dtcp} and the proposed co-design with task-oriented \gls{jscc}. The numerical and experimental results are provided in \cref{sec_result}. Finally, conclusions are drawn in \cref{sec_conclusion}.

The main notations used throughout the paper are summarized in \cref{tab_notation}. To improve readability and manage the complexity of the joint design of communication, computation, and control, the temporal subscript of the notation is omitted in \cref{sec_system}.

\input{Table/tab_notation}

%% file: Table/tab_notation.tex
\begin{table*}[!htbp]
\centering
\footnotesize
\caption{SUMMARY OF MAIN SYMBOLS}
\label{tab_notation}
\renewcommand{\arraystretch}{1.3}
\begin{tabular}{|p{55pt}|p{165pt}|!{\vrule width 0.1pt}p{55pt}|p{165pt}|}
\hline
$\bm{x}$                 & Input image (data)                                                           & $K_b$                                                                                                                                                     & Size of mini-batch                                                      \\ \hline
$X$                      & Random variable of $\bm{x}$                                                  & $\bm{x}_t$                                                                                                                                                & Image captured by a camera at time~$t$                                  \\ \hline
$\mathcal{X}$            & Space of $\bm{x}$                                                            & $\bm{z}_{t}^{l(t)}, \bm{z}_{t}^{l}$                                                                                                                       & JSCC symbols with length~$l$~transmitted at time~$t$~                   \\ \hline
$\bm{z}$                 & Transmitted JSCC symbols                                                     & $\hat{\bm{z}}_{t-\delta_{eu}}^{l(t-\delta_{eu})},\newline \hat{\bm{z}}_{t-\delta_{eu}}^{l}$                                                               & Reconstructed JSCC symbols with length~$l$~received at time~$t$         \\ \hline
$Z$                      & Random variable of $\bm{z}$                                                  & $\bm{m}_t$                                                                                                                                                & State information transmitted at time~$t$                               \\ \hline
$\hat{\bm{z}}$           & Reconstructed JSCC symbols                                                   & $\tau$                                                                                                                                                    & Duration of time slot                                                   \\ \hline
$\mathcal{Z}$            & Space of JSCC symbols                                                        & $D_t$                                                                                                                                                     & Index set for~$\bm{z}_{t}^{l}$                                          \\ \hline
$\check{\bm{z}}$         & Received JSCC symbols                                                        & $\delta_e$                                                                                                                                                & Computing delay of JSCC encoding                                        \\ \hline
$h$                      & Channel state                                                                & $\delta_u$                                                                                                                                                & Uplink communication delay                                              \\ \hline
$\bm{n}$                 & Gaussian noise                                                               & $\delta_a$                                                                                                                                                & Computing delay of agent                                                \\ \hline
$H$                      & Random variable of $h$                                                       & $\delta_d$                                                                                                                                                & Downlink communication delay                                            \\ \hline
$\mathcal{H}$            & Space of $h$                                                                 & $\delta_c$                                                                                                                                                & Control delay                                                           \\ \hline
$\bm{m}$                 & State information                                                            & $\delta_{eu}$                                                                                                                                             & Combined delay of $\delta_e$~and~$\delta_u$                             \\ \hline
$M$                      & Random variable of $\bm{m}$                                                  & $\delta$                                                                                                                                                  & End-to-end delay                                                        \\ \hline
$\mathcal{M}$            & Space of $\bm{m}$                                                            & $\delta_{T}$                                                                                                                                              & Delay threshold                                                         \\ \hline
$\bm{a}$                 & Ground-truth action                                                          & $l_p$                                                                                                                                                     & Prediction horizon                                                      \\ \hline
$A$                      & Random variable of $\bm{a}$                                                  & $l_w$                                                                                                                                                     & Extra Prediction Horizon                                                \\ \hline
$\hat{\bm{a}}$           & Estimated action                                                             & $\bm{r}_{t-\delta_{eu}}^{\text{traj}}$                                                                                                                    & Trajectory feature corresponding to~$x_{t-\delta_{eu}}$                 \\ \hline
$\hat{A}$                & Random variable of $\hat{\bm{a}}$                                            & $\mathcal{R}_\text{traj}$                                                                                                                                 & Space of trajectory features                                            \\ \hline
$\mathcal{A}$            & Space of action                                                              & $\bm{h}_{t-\delta_{eu}}^{\text{traj}}$                                                                                                                    & Trajectory hidden state~corresponding to~$x_{t-\delta_{eu}}$            \\ \hline
$f_e(\cdot)$             & Function of JSCC encoder                                                     & $\bm{w}_{t-\delta_{eu}}$                                                                                                                                  & Planned waypoint at time~$t-\delta_{eu}$                                \\ \hline
$f_h(\cdot)$             & Function of noisy fading channel                                             & $\bm{c}_{t-\delta_{eu}+l_p}^{\text{traj}}$                                                                                                                & Predicted trajectory command with horizon $l_p$ for~$x_{t-\delta_{eu}}$ \\ \hline
$f_a(\cdot)$             & Function of autonomous driving agent                                         & $\bm{r}_{t-\delta_{eu}}^{\text{ctrl}}$                                                                                                                    & Control feature corresponding to~$x_{t-\delta_{eu}}$                    \\ \hline
$P_{\text{target}}$      & Average power constraint of~$\bm{z}$                                         & $\mathcal{R}_\text{ctrl}$                                                                                                                                 & Space of control features                                               \\ \hline
$l_x$                    & Dimension of~$\bm{x}$                                                        & $\bm{h}_{t-\delta_{eu}}^{\text{ctrl}}$                                                                                                                    & Control hidden state~corresponding to~$x_{t-\delta_{eu}}$               \\ \hline
$l_z$                    & Dimension of~$\bm{z}$                                                        & $\bm{c}_{t-\delta_{eu}+l_p}^{\text{ctrl}}$                                                                                                                & Predicted control command with horizon $l_p$ for~$x_{t-\delta_{eu}}$    \\ \hline
$g(\cdot, \cdot)$        & Distortion measuring function                                                & $\bm{c}_{t-\delta_{eu}+l_p}^{\text{comb}}$                                                                                                                & Predicted combined command with horizon $l_p$ for~$x_{t-\delta_{eu}}$   \\ \hline
$\zeta_{\text{ratio}}$   & Constraint of bandwidth compression ratio                                    & $\mathcal{C}_\text{cmd}$                                                                                                                                  & Space of commands                                                       \\ \hline
$\zeta_{\text{rate}}$    & Constraint of rate                                                           & $f_\text{feat-t}(\cdot, \cdot, \cdot)$                                                                                                                    & Function of trajectory feature extractor                                \\ \hline
$\phi$                   & Parameters of JSCC encoder                                                   & $f_\text{traj}(\cdot)$                                                                                                                                    & Function of trajactory branch                                           \\ \hline
$\psi$                   & Parameters of autonomous driving agent                                       & $f_\text{feat-c}(\cdot, \cdot, \cdot)$                                                                                                                    & Function of control feature extractor                                   \\ \hline
$\beta$                  & Lagrange multiplier                                                          & $f_\text{ctrl}(\cdot, \cdot)$                                                                                                                             & Function of control branch                                              \\ \hline
$f_{\bm{\mu}}(\cdot)$    & Function for estimating the mean of reconstructed JSCC symbols               & $f_\text{comb}(\cdot, \cdot)$                                                                                                                             & Function of command combination                                         \\ \hline
$f_{\bm{\sigma}}(\cdot)$ & Function for estimating the standard deviation of reconstructed JSCC symbols & $\lambda_c, \lambda_\text{feat}, \lambda_\text{value},\newline \lambda_\text{speed},\lambda_\text{traj}, \lambda_\text{ctrl},\newline \lambda_\text{aux}$ & Hyperparameters of agent                                                \\ \hline
$K_a$                    & Size of dataset                                                              & $i$                                                                                                                                                    & General index depended on the context                                   \\ \hline
\end{tabular}
\end{table*}

%% file: 1_2_related_work.tex
\section{Related Works}
\label{sec_related_works}
\paragraph{Task-oriented Joint Source-Channel Coding}
The integration of deep learning into communication systems has introduced groundbreaking possibilities that go beyond conventional boundaries \cite{Guenduez_2023_BTB}. Unlike traditional methods that separate source and channel coding, deep \gls{jscc} offers an end-to-end learning-based approach, optimizing the information transmission holistically. Previous research has highlighted the superiority of deep \gls{jscc} for reconstruction-oriented communication \cite{Kurka_2019_SRo, Bourtsoulatze_2019_DJS, Kurka_2020_DfD} over traditional source coding (e.g., JPEG \cite{Wallace_1992_TJs} and JPEG2000 \cite{Taubman_2002_JIc}) and channel coding (e.g., LDPC codes \cite{Gallager_1962_Ldp}) methods, especially in low-\gls{snr} environments.

However, most reconstruction-oriented approaches have focused primarily on data-centric metrics, often leading to suboptimal task performance. High-fidelity reconstructions are not always necessary for machine-to-machine communication, whereas preserving task-specific information is more important \cite{Diao_2024_TOS, Diao_2024_TTG, Chaccour_2024_LDM, Yang_2023_SCf, Strinati_2024_GOa, Pandey_2023_GOC, Kang_2023_PSi}. Driven by the growing demand for task-oriented communication designs, researchers have increasingly explored the \gls{ib} approach. The \gls{ib} \cite{Tishby_1999_TIB} method seeks to maximize the preservation of task-specific information while minimizing task-agnostic information from input. Traditional \gls{ib} approach relied on the computationally intensive Blahut-Arimoto algorithm \cite{Arimoto_1972_Aaf, Blahut_1972_Coc}, which was impractical for deep learning applications due to its complexity \cite{Tishby_2015_Dla}. This limitation was addressed by the introduction of a variational approach to the \gls{ib} method, known as \gls{vib} \cite{Alemi_2017_DVI}, which made it feasible to apply \gls{ib} principles in deep learning by approximating the true posterior with a variational distribution.

Recent studies have successfully integrated \gls{vib} with deep \gls{jscc}, formalizing task-oriented communication strategies that outperform traditional reconstruction-oriented frameworks. For example, recent works \cite{Shao_2022_LTO, Shao_2023_TOC} have demonstrated that combining \gls{vib} with deep \gls{jscc} can significantly improve communication efficiency and robustness, particularly in scenarios where it is essential to prioritize task-specific information over the fidelity of raw data. Another study \cite{Liao_2024_AAG} focused on applying semantic communication for camera relocalization, optimizing the trade-off between inference accuracy and \gls{e2e} latency.

\paragraph{Edge Inference}
To meet the stringent latency demands of modern applications, edge inference has emerged as a crucial solution that effectively addresses the limitations of traditional cloud-based architectures \cite{Shi_2020_CEE, Li_2018_EIO}. A key innovation in this domain is the \textit{split inference}, where the inference process is divided between the edge device and the server. In split inference architectures, a lightweight neural network on the mobile device extracts compact feature vectors from the data, which are then transmitted to an edge server for further analysis. This architecture allows for initial data processing on the device, followed by more intensive computations at the edge, thus reducing latency and improving overall system efficiency \cite{Huang_2020_DCR, Shi_2019_IDE, Li_2020_EAO, Shao_2020_CCT, Shao_2020_BAE, Jankowski_2020_JDE, Jankowski_2021_WIR}. Deep \gls{jscc} plays a critical role in this process by enabling efficient compression and transmission of these intermediate features \cite{Shao_2020_CCT, Shao_2020_BAE, Jankowski_2020_JDE, Jankowski_2021_WIR, Shao_2022_LTO, Shao_2023_TOC}. 

Recent studies \cite{Dubois_2021_LCf, Shao_2021_BGA, Shao_2022_LTO, Shao_2020_BAE, Shao_2023_TOC} have increasingly focused on inference accuracy as a key performance indicator, highlighting the shift in communication systems towards optimizing task-specific performance rather than data-centric metrics. In particular, \cite{Shao_2022_LTO} introduced a method that dynamically adjusts the length of the transmitted signal in response to varying communication conditions, ensuring that required inference accuracy is maintained while optimizing the use of available resources.

\paragraph{Predictions in URLLC Applications}
In the context of \gls{urllc}, prediction-based methods have been explored to mitigate delays. For instance, \cite{Simsek_2016_5ET} proposed a technique to predict movements or force feedback to reduce perceived delay in Tactile Internet applications. Similarly, \cite{Tong_2018_MWR} presented a co-design approach for packetized predictive control (PPC) in real-time \gls{cps}, addressing the delay in the tight interaction between wireless communication and control systems. For visual content, 
\cite{Richter_2019_ARP} investigated how predictive displays can mitigate communication delays in telesurgery using \gls{ar} technology. The proposed system provided real-time visual feedback to surgeons by predicting movements of robotic tools, significantly improving task completion times under latency without increasing error rates.
Likewise, \cite{Hou_2020_MPa} introduced edge intelligence to predict user motion, enabling pre-rendering and caching of \gls{vr} content, thus significantly reducing the latency in \gls{vr} streaming. 
However, in these studies \cite{Simsek_2016_5ET, Tong_2018_MWR, Richter_2019_ARP, Hou_2020_MPa}, the trade-off between the prediction horizon and the reliability of the system was not adequately addressed.
To bridge this gap, \cite{Hou_2020_PaC} focused on the challenges of delay and reliability in \gls{urllc} by co-designing prediction and communication systems. 
The proposed framework enables mobile devices to predict future states and send these predictions to a data center in advance, thus reducing perceived delays. The study also analyzed the trade-off between prediction accuracy and system reliability, demonstrating that longer prediction horizons increase the likelihood of errors.

%% file: 2_1_IB.tex
\section{System Model and Problem Formulation}
\label{sec_system}
\begin{figure*}[t]
    \begin{center}
    \includegraphics[width=0.95\linewidth]{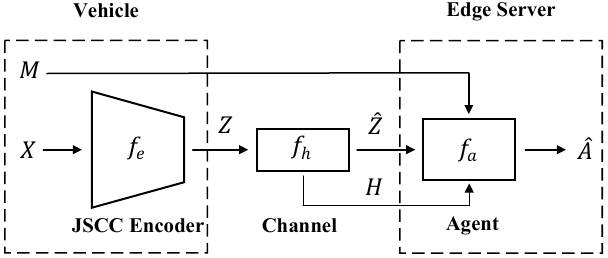}
    \end{center}
       \caption{General framework of edge-enabled autonomous driving.}
    \label{fig_IB_framework}
\end{figure*}
As shown in \cref{fig_IB_framework}, we consider an edge server that provides computing service for a single vehicle (device). The vehicle transmits \gls{jscc} symbols and encoded state information to the edge server. After processing the received data, the edge server sends the drive commands back to the vehicle. 

The on-vehicle \gls{jscc} encoder \( 
f_e \) is defined as: 
\begin{align}
    f_e: \mathcal{X} \rightarrow \mathcal{Z}: \bm{x} \mapsto \bm{z},
    \label{eq_JSCC_encoder}
\end{align}
where $\bm{x} \in \mathbb{R}^{l_x}$ denotes the input image, and $\bm{z} = [z_1, \dots, z_{l_z}] \in \mathbb{C}^{l_z}$ denotes the transmitted \gls{jscc} symbols. Here, $l_x$ denotes the source bandwidth, which is the product of the height, width, and number of color channels of the image \(\bm{x}\). The parameter $l_z$ denotes the channel bandwidth. We define $l_z/l_x$ as the \textit{bandwidth compression ratio} \cite{Bourtsoulatze_2019_DJS}. In particular, the transmitted \gls{jscc} symbols should satisfy the average power constraint \(P_{\text{target}}\):
\begin{align}
    \frac{1}{l_z}\sum_{i=1}^{l_z} |z_i|^2 \leq P_{\text{target}}.
    \label{eq_power_constraint}
\end{align}

Then $\bm{z}$ are transmitted to the edge server via communication channels, which can be mathematically represented by the function:
\begin{align}
    f_h: \mathcal{Z} \rightarrow \mathcal{Z}: \bm{z} \mapsto \hat{\bm{z}}.
    \label{eq_JSCC_channel}
\end{align}

In this paper, we model the communication channel as a frequency-selective channel implemented through \gls{ofdm} to mitigate multipath fading, as detailed in \cite{Yang_2022_OGD}:
\begin{align}
    \check{\bm{z}} = f_h(\bm{z}) = \bm{h} \cdot \bm{z} + \bm{n},
    \label{eq_ofdm_channel}
\end{align}
where $\check{\bm{z}}$ denotes the received \gls{jscc} symbols, and $\bm{n} \sim \mathcal{CN}(0, \bm{\sigma}_{n}^{2} I)$ represents complex Gaussian noise with zero mean and standard deviation $\bm{\sigma}_{n}$, where \( I \) denotes the identity matrix and \(\bm{\sigma}_{n}\) is a diagonal matrix. The channel frequency response $\bm{h}\in \mathbb{C}^{l_z}$ captures the characteristics of multipath fading. A comprehensive modeling of the OFDM channel is provided in Appendix \ref{apd_freq_selec}.

In this paper, we assume the perfect \gls{csi}. After receiving, the \gls{jscc} symbols are equalized:
\begin{align}
    \hat{\bm{z}} = \frac{h^*}{|h|^2}\check{\bm{z}},
\end{align}
where $h^*$ denotes the conjugate of channel coefficient $h$ and $\hat{\bm{z}}$ denotes the reconstructed \gls{jscc} symbols.

Following transmission, the reconstructed \gls{jscc} symbols $\hat{\bm{z}}$ are processed by the autonomous driving agent $f_a$ with state information and channel state, which is defined as:
\begin{align}
    f_a: \mathcal{Z}\times\mathcal{M}\times\mathcal{H} \rightarrow \mathcal{A}: (\hat{\bm{z}}, \bm{m}, \bm{h}) \mapsto \hat{\bm{a}},
    \label{eq_agent}
\end{align}
where $\hat{\bm{a}} \sim p(\hat{\bm{a}})$ denotes the estimated action, which approximates the ground-truth action $\bm{a} \sim p(\bm{a})$. In addition, $\bm{m}$ denotes state information consisting of vehicle speed, discrete navigation command, destination coordinates, and timestamp. The agent incorporates vehicle state information $\bm{m}$ and channel state information $\bm{h}$ as conditional inputs, establishing a direct link between communication and control to improve decision-making. Since the state information $\bm{m}$ typically consumes negligible bandwidth (in this work, $\bm{m}$ consists of four floating-point numbers and one integer), we assume that it is received losslessly by the edge server, provided that the corresponding image is successfully received and decoded.

The task-oriented objective of edge-enabled autonomous driving is to minimize the expected distortion between the ground-truth action \(\bm{a}\) and the estimated action \(\hat{\bm{a}}\), which is defined as \( g:\mathcal{A}\times\mathcal{A} \rightarrow \mathbb{R}^{+} \). Consequently, the problem of the proposed task-oriented co-design is defined in \cref{pro_distortion}.

\begin{problem} 
    \begin{align}
        \min_{f_e, f_a} \ &\mathbb{E}_{\bm{a}\sim p(\bm{a}|\hat{\bm{z}},\bm{m},\bm{h})}\left\{\mathbb{E}_{\hat{\bm{a}}\sim p(\hat{\bm{a}}|\hat{\bm{z}},\bm{m},\bm{h})} \left[g(\bm{a},\hat{\bm{a}})\right]  \right\} \label{eq_main_problem} \\
        \textnormal{s.t.} \quad&l_z/l_x - \zeta_{\textnormal{ratio}} \leq 0, \label{eq_main_problem_constraints}  \\
        & (\ref{eq_JSCC_encoder}),
        (\ref{eq_power_constraint}),
        (\ref{eq_JSCC_channel}), (\ref{eq_ofdm_channel}), (\ref{eq_agent}),
    \end{align} 
    where \(\zeta_{\textnormal{ratio}}\) denotes an upper bound of the bandwidth compression ratio.
    \label{pro_distortion}
\end{problem}

\cref{pro_distortion} is an abstract formulation that describes the overarching goal of optimizing task performance through appropriate parameter selection. However, directly solving \cref{pro_distortion} poses significant computational challenges, particularly in the evaluation of the expectation over random variables, which involves integration that can be computationally prohibitive. In addition, finding a proper objective function $g(\cdot,\cdot)$ is also difficult. In the following section, we introduce the \gls{vib} approach combined with \glspl{dnn} to effectively address \cref{pro_distortion}.

\section{Variational Information Bottleneck Approach}
\label{sec_VIB}
\begin{figure}
    \begin{center}
    \begin{tikzpicture}[node distance=0.5cm, every node/.style={font=\large, baseline}]
        \node (A) at (0,0) {$A$};
        \node (X) [right=of A] {$X$};
        \node (Z) [right=of X] {$Z$};
        \node (Z_hat) [right=of Z] {$\hat{Z}$};
        \node (A_hat) [right=of Z_hat] {$\hat{A}$,};
        \node (M) [below=of Z_hat] {$M$};
        \node (H) [above=of Z_hat] {$H$};
        
        \draw[->] (A) -- (X);
        \draw[->] (X) -- (Z);
        \draw[->] (Z) -- (Z_hat);
        \draw[->] (Z_hat) -- (A_hat);
        \draw[->] (M) -- (A_hat);
        \draw[->] (H) -- (A_hat);
        \draw[->] (A) -- (H);
        \draw[->] (A) -- (M);
    \end{tikzpicture}
    \end{center}
    \caption{The DPGM for edge-enabled autonomous driving.}
    \label{fig_DPGM}
\end{figure}
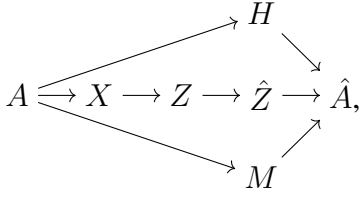
Based on the discussion in \cref{sec_system}, the \gls{dpgm} of the proposed framework can be depicted as shown in \cref{fig_DPGM}, where \(A\), \(X\), \(Z\), \(\hat{Z}\), \(M\), \(H\) and \(\hat{A}\) denotes the random variables of the ground-truth action \(\bm{a}\), input images \(\bm{x}\), transmitted \gls{jscc} symbols \(\bm{z}\), reconstructed \gls{jscc} symbols \(\hat{\bm{z}}\), state information \(\bm{m}\), channel state \(\bm{h}\), and estimated action \(\hat{\bm{a}}\), respectively. With this \gls{dpgm}, the \gls{ib} can be formulated as an optimization problem:
\begin{align}
    \min_{\phi, \psi} \quad&-I(A;\hat{Z}, M, H) \notag\\
    \text{s.t.} \quad&I(X;\hat{Z})-\zeta_{\text{rate}} \leq 0,
    \label{eq_IB_optim}
\end{align}
where \(\phi\) and \(\psi\) are the parameters of \gls{jscc} encoder \(f_e\) and autonomous driving agent \(f_a\), respectively. \( \zeta_{\text{rate}} \) denotes the upper bound of the rate. \cref{eq_IB_optim} is derived as a practical and task-oriented reformulation of \cref{pro_distortion}. The objective function in \cref{eq_IB_optim}, $I(A;\hat{Z}, M, H)$, measures the information shared between the target variable $A$ and the combined set of variables $(\hat{Z}, M, H)$. By minimizing $-I(A;\hat{Z}, M, H)$, we ensure that the information about $A$ retained in $(\hat{Z}, M, H)$ is maximized, aligning the optimization with the task-specific objectives of \cref{pro_distortion}. This is supported by the \gls{dpgm} structure where $\hat{A}$ depends on $(\hat{Z}, M, H)$, highlighting that preserving information about $A$ in these variables is key to achieving optimal task performance. Moreover, the constraints in \cref{eq_IB_optim} explicitly incorporate communication limitations through the term $I(X,\hat{Z})$, which limits the rate of transmitted information. These constraints parallel the bandwidth restrictions in \cref{pro_distortion}, thereby ensuring consistency between the two formulations.

By introducing the Lagrange multiplier \( \beta \), \gls{ib} can be further formulated to minimize the following objective function: 
\begin{align}
    \mathcal{L}_{\text{IB}}:= \underbrace{-I(A;\hat{Z}, M, H)}_{\text{Distortion}} + \beta \underbrace{I(X;\hat{Z})}_{\text{Rate}}.
    \label{eq_IB_Lagrange}
\end{align}

Building on the foundational works of \cite{Shao_2022_LTO} and \cite{Alemi_2017_DVI}, we develop a \gls{vib} approach to approximate each term in \cref{eq_IB_Lagrange}, addressing the intractability of mutual information. The first term \( -I(A;\hat{Z},M,H) \) can be expressed as:
\begin{align}
    -I(A;&\hat{Z},M,H) \notag\\
    &= -\int p(\bm{a},\hat{\bm{z}}, \bm{m}, \bm{h})\log{\frac{p(\bm{a}|\hat{\bm{z}}, \bm{m}, \bm{h})}{p(\bm{a})}} \dif \bm{a} \dif \hat{\bm{z}} \dif \bm{m} \dif \bm{h}\notag\\
    &= -\int p(\bm{a},\hat{\bm{z}}, \bm{m}, \bm{h})\log{p(\bm{a}|\hat{\bm{z}},\bm{m},\bm{h})} \dif \bm{a} \dif \hat{\bm{z}} \dif \bm{m} \dif \bm{h} \notag\\
    &\quad- H(A),
\end{align}
where \(H(A)\) denotes the entropy of random variable \(A\), which is independent of the optimization and thus can be ignored. In addition, \( p(\bm{a}|\hat{\bm{z}}, \bm{m}, \bm{h}) \) is the posterior probability, which can be derived from the \gls{dpgm} \cite{Alemi_2017_DVI, Shao_2022_LTO} as:
\begin{align}
    &p(\bm{a}|\hat{\bm{z}}, \bm{m}, \bm{h}) = \int p(\bm{a},\bm{x},\bm{z}|\hat{\bm{z}},\bm{m},\bm{h})\dif{\bm{x}}\dif{\bm{z}} \notag\\
    &= \int\frac{p(\bm{a})p(\bm{x}|\bm{a})p_{\phi}(\bm{z}|\bm{x})p(\hat{\bm{z}}|\bm{z})p(\bm{m}|\bm{a})p(\bm{h}|\bm{a})}{p(\hat{\bm{z}}, \bm{m},\bm{h})} \dif{\bm{x}}\dif{\bm{z}}.
\end{align}
Since this integration is intractable in our case, we use \( q_{\psi}(\bm{a}|\hat{\bm{z}},\bm{m}, \bm{h}) \) as a variational approximation of \(p(\bm{a}|\hat{\bm{z}},\bm{m},\bm{h})\). Based on the fact that \gls{kl} divergence is always non-negative, the following inequality can be obtained:
\begin{align}
    -I(A;&\hat{Z}, M, H) + H(A) \notag\\
    &=\mathbb{E}_{\bm{a},\bm{x}}\biggl[\mathbb{E}_{\hat{\bm{z}}|\bm{x};\phi}\bigl[\mathbb{E}_{\bm{m},\bm{h}}[-\log p(\bm{a}|\hat{\bm{z}},\bm{m},\bm{h})]\bigr]\biggr] \notag\\
&\leq \mathbb{E}_{\bm{a},\bm{x}}\biggl[\mathbb{E}_{\hat{\bm{z}}|\bm{x};\phi}\bigl[\mathbb{E}_{\bm{m},\bm{h}}[-\log q_{\psi}(\bm{a}|\hat{\bm{z}},\bm{m},\bm{h})]\bigr]\biggr].
\label{eq_expectation_inequal}
\end{align}

The second term \( I(X;\hat{Z}) \) can be formulated as:
\begin{align}
    I(X;\hat{Z})=\mathbb{E}_{\bm{a},\bm{x}}\left[D_{\text{KL}}(p_{\phi}(\hat{\bm{z}} | \bm{x}) \| p(\hat{\bm{z}})) \right],
\end{align}
where \( p(\hat{\bm{z}}) \) is the intractable prior probability of \( \hat{\bm{z}} \). Instead of using approximation proposed in \cite{Molchanov_2017_VDS}, we adopt a predefined Gaussian distribution \(q(\hat{\bm{z}}) \sim \mathcal{N}(\bm{\mu}_{\hat{\bm{z}}}, \bm{\sigma}^2_{\hat{\bm{z}}}I)\) as the approximation of \( p(\hat{\bm{z}}) \) \cite{Kingma_2013_Aev}, where $\bm{\mu}_{\hat{\bm{z}}}$ and $\bm{\sigma}_{\hat{\bm{z}}}$ represent the mean and standard deviation of the Gaussian distribution, respectively. 

In addition, we model the \gls{jscc} encoder \( f_e \) as a probability model \( p_{\phi}(\bm{z}|\bm{x})\). Considering that \( p_{\phi}(\hat{\bm{z}}|\bm{x}) = \int p_{\phi}(\bm{z}|\bm{x})p(\hat{\bm{z}}|\bm{z}) \dif \bm{z} \), where \(p(\hat{\bm{z}}|\bm{z})\) represents the probabilistic nature of the channel function \(f_h\), and assuming perfect \gls{csi}, we define \( p_{\phi}(\hat{\bm{z}}|\bm{x}) \sim \mathcal{N}(f_{\bm{\mu}}(\hat{\bm{z}}), f_{\bm{\sigma}}^2(\hat{\bm{z}})I) \). Here, $f_{\bm{\mu}}(\cdot)$ and $f_{\bm{\sigma}}(\cdot)$ are functions that estimate the mean and standard deviation of this Gaussian distribution, respectively.

Since \gls{kl} divergence is always non-negative, we have \( D_{\text{KL}}(p(\hat{\bm{z}})\parallel q(\hat{\bm{z}})) \geq 0 \), which can derive the upper bound of \( I(X;\hat{Z}) \):
\begin{align}
    I(X;\hat{Z}) \leq \mathbb{E}_{\bm{a},\bm{x}}\left[D_{\text{KL}}(p_{\phi}(\hat{\bm{z}} | \bm{x}) \| q(\hat{\bm{z}})) \right]. 
    \label{eq_upper_bound_2nd}
\end{align}
The derivation of \cref{eq_upper_bound_2nd} is given in Appendix \ref{apd_upper_bound_2nd}. Therefore, we derive the following corollary as an approximation of \cref{eq_IB_Lagrange}.
\begin{corollary}
    Assume the \gls{dpgm} shown in \cref{fig_DPGM}, let \( q_{\psi}(\bm{a}|\hat{\bm{z}},\bm{m},\bm{h}) \) be a variational approximation of \( p(\bm{a}|\hat{\bm{z}},\bm{m},\bm{h}) \), let \(q(\hat{\bm{z}}) \sim \mathcal{N}(\bm{\mu}_{\hat{\bm{z}}}, \bm{\sigma}^2_{\hat{\bm{z}}}I)\) be a variational approximation of \( p(\hat{\bm{z}}) \), and let \( p_{\phi}(\hat{\bm{z}}|\bm{x}) \sim \mathcal{N}(f_{\bm{\mu}}(\hat{\bm{z}}), f_{\bm{\sigma}}^2(\hat{\bm{z}})I) \) be a variational approximation of \( p(\hat{\bm{z}}|\bm{x}) \), the upper bound of \cref{eq_IB_Lagrange} is given by
    \begin{align}
        \mathcal{L}_{\textnormal{VIB}}:=&\ \mathbb{E}_{\bm{a},\bm{x}}\Bigl\{
        \mathbb{E}_{\hat{\bm{z}}|\bm{x};\phi}\bigl[\mathbb{E}_{\bm{m},\bm{h}}\left[-\log q_{\psi}(\bm{a}|\hat{\bm{z}},\bm{m},\bm{h})\right]\bigr] \notag\\
        &\qquad \ \ +\beta D_{\textnormal{KL}}(p_{\phi}(\hat{\bm{z}} | \bm{x}) \| q(\hat{\bm{z}}))
        \Bigr\} \notag\\
         \geq&\ \mathcal{L}_\textnormal{IB}+H(A).
    \end{align}
\end{corollary}

This corollary can be optimized using stochastic gradient descent through Monte Carlo sampling, providing a practical framework for empirical estimation and subsequent optimization. 
Given a dataset with size \(K_a\), a mini-batch $\{(\bm{a}_i,\bm{x}_i)\}^{K_{b}}_{i=1}$ of size \(K_b\) is randomly drawn without overlap in the same epoch to compute the gradient of loss \(\mathcal{L}_\text{VIB}\). In particular, the number of samples of \(-\log q_{\psi}(\bm{a}|\hat{\bm{z}},\bm{m},\bm{h})\) can be set to 1 as long as the size of the dataset \(K_a\) is large enough \cite{Kingma_2013_Aev}. Thus, we have the following estimation:
\begin{align}
    \mathcal{L}_{\text{VIB}}\approx \frac{1}{K_{b}}\sum_{i=1}^{K_{b}} 
\bigl\{&
-\log q_{\psi}(\bm{a}_{i}|\hat{\bm{z}}_{i},\bm{m}_{i}, \bm{h}_i)  \notag\\ 
&+\beta D_{\text{KL}}(p_{\phi}(\hat{\bm{z}} | \bm{x}_{i}) \| q(\hat{\bm{z}}))
\bigr\}.
\label{eq_VIB_realization}
\end{align}
Note that the dataset $\{(\bm{a}_i,\bm{x}_i)\}^{K_{a}}_{i=1}$ can be collected from expert agents or human drivers.

%% file: 2_2_DTCP.tex
\section{Delay-Aware Trajectory-Guided Control Prediction}
\label{sec_DTCP}
\begin{figure*}[t]
    \begin{center}
    \includegraphics[width=0.95\linewidth]{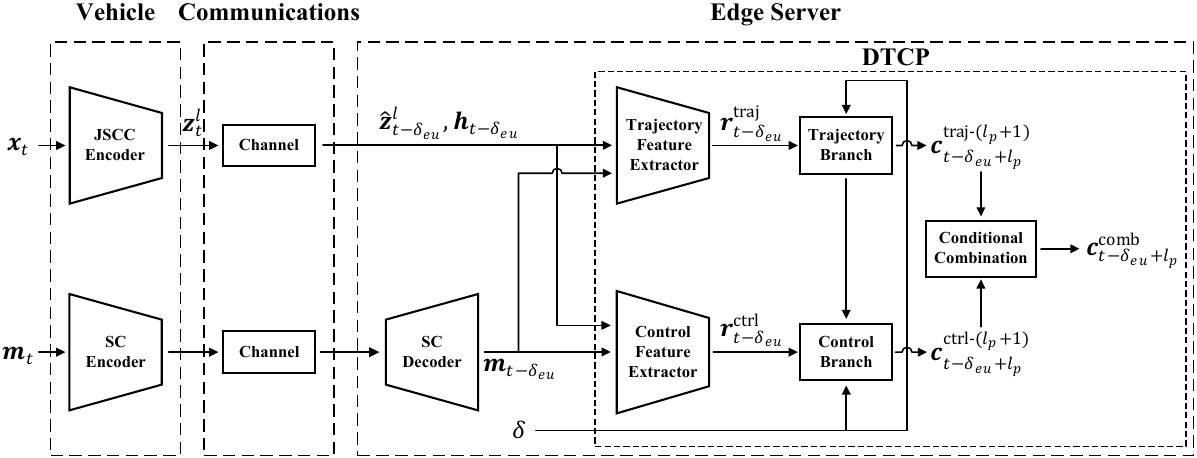}
    \end{center}
       \caption{The proposed task-oriented co-design framework based on JSCC and DTCP.}
    \label{fig_general_framework}
\end{figure*}
\gls{tgcp} is one of the state-of-the-art frameworks of \gls{e2e} autonomous driving, integrating trajectory planning and multistage control prediction together \cite{Wu_2022_TgC}. This advanced framework, which uses only a monocular camera, currently ranks third on the \gls{carla} leaderboard\footnote{\url{https://leaderboard.carla.org/leaderboard/}}. However, the original \gls{tgcp} framework relies on a stream of raw images for decision making, resulting in high bandwidth usage. In addition, it does not account for the impact of communication latency on decision-making processes.

\glsreset{dtcp}
To overcome these limitations, we have developed a \gls{dtcp} strategy that integrates the trajectory and control branches while considering the delay, as shown in \cref{fig_general_framework}. This integration ensures that predicted drive commands reduce the perceived \gls{e2e} delay, leading to safer and more efficient autonomous driving.

We assume the system is time-slotted and initiates at the time \(t=0\). The duration of each time slot is denoted as \(\tau\). 
Let \(\bm{z}_{t}^{l(t)}\) denote the transmitted \gls{jscc} symbols with length \(l(t)\), and \(\bm{m}_{t}\) represents the state information, both corresponding to the image \(\bm{x}_{t}\) captured by the onboard camera at time $t$. The term \(1\leq l(t)\leq l_z\) denotes a function that decides the number of selected \gls{jscc} symbols of \(\bm{z}_t\). For simplicity of notation, we denote \(l(t)\) by \(l\) in the following formulation.

Define the index set \(D_{t} \subset \{1,2,\dots,l_z\} \) such that:
\begin{align}
    D_{t} = \{&i\in\{1,\dots,l_z\} \mid \notag\\
    &|z_i^t|^2 \text{ is one of the } l \text{ largest numbers in } |\bm{z}_t|^2\}.
\end{align} 
In particular, \(\bm{z}_{t}^{l}=\{z^{t}_{i}\mid i\in D_{t}\}\) are \gls{jscc} symbols selected from \(\bm{z}_{t} = [z_1^{t}, \dots, z_{l_z}^{t}]\) based on energy significance. The selected \gls{jscc} symbols are kept for transmission, while the missing \gls{jscc} symbols are filled with 0 on the edge server. Note that this selection process can be integrated into the \gls{jscc} encoder \(f_e\). If only selected \gls{jscc} symbols are transmitted, the receiver must be made aware of the indices of the selected or abandoned \gls{jscc} symbols, which may increase the communication load. In this work, we design the selective \gls{jscc} symbols to provide flexibility within this task-oriented communication co-designed paradigm and demonstrate their potential for dimensionality reduction. To address the additional communication load introduced by index transmission, methods such as Variable-Length Variational Feature Encoding (VL-VFE) \cite{Shao_2022_LTO} offer promising directions for further exploration.

\subsection{Prediction for End-to-End Delay}
\label{subsec_pred_e2e_delay}
\begin{figure}[t]
    \begin{center}
    \includegraphics[width=0.95\linewidth]{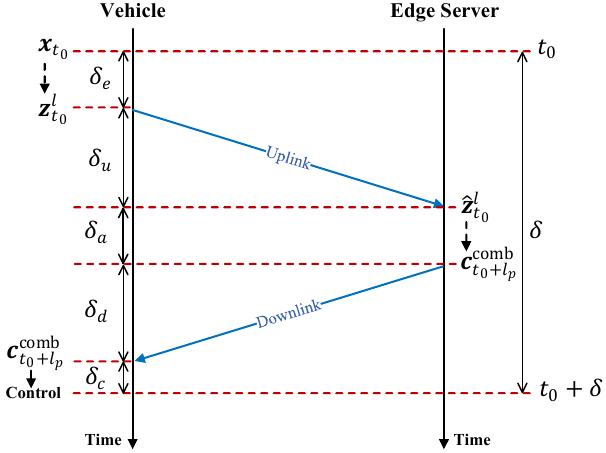}
    \end{center}
       \caption{The illustration of a completed cycle of the communication, computing, and control process, along with the prediction structure.}
    \label{fig_prediction_structure}
\end{figure}

In edge-enabled autonomous systems, drive commands are often outdated due to \gls{e2e} delay, including communication, computation, and control delays. \cref{fig_prediction_structure} shows a complete cycle of the communication, computing, and control process, along with the prediction structure. Assume that an image \(\bm{x}_{t_0}\) is captured by the camera at time \(t=t_0\). After encoding and selecting, the \gls{jscc} symbols \(\bm{z}_{t_0}^{l}\) are generated with a computation delay \(\delta_{e}\). The \gls{jscc} symbols reconstructed by the edge server are denoted as \(\hat{\bm{z}}_{t_0}^{l}\) arriving with an uplink delay \(\delta_{u}\). The agent on the edge server takes \(\delta_a\) time slots to generate the command \(\bm{c}_{t_0+l_p}^{\text{comb}}\), where \(l_p \geq 0\) denotes the prediction horizon. The command is then sent back to the vehicle with a downlink delay \(\delta_d\). Upon receiving the command, the vehicle takes \(\delta_c\) time slots to execute the command. Thus, the \gls{e2e} delay is expressed as \(\delta = \delta_e + \delta_u + \delta_a + \delta_d + \delta_c\). Consequently, the perceived \gls{e2e} delay is given by \(\delta - l_p\). Since the command \(\bm{c}_{t_0+l_p}^{\text{comb}}\) consumes negligible bandwidth, it is assumed to be transmitted losslessly to the vehicle in this work. It is worth noting that while the uplink delay can be further decomposed into components such as transmission delay, queuing delay, propagation delay, and processing delay, a detailed analysis of each individual component lies outside the primary scope of this work. Instead, our focus is on the \gls{e2e} delay and addressing it through predictive mechanisms.

It is crucial to recognize that in the process described above, when the onboard camera captures an image \(\bm{x}_t\) at any time \(t\), reconstructed \gls{jscc} symbols \(\hat{\bm{z}}_{t-\delta_{eu}}^{l}\) (corresponding to image \(\bm{x}_{t-\delta_{eu}}\)) on the edge server are outdated of \(\delta_{eu}=\delta_e+\delta_u\) time slots. The combined delay \(\delta_{eu}\) can be calculated from the timestamp in the state information \(\bm{m}_{t-\delta_{eu}}\), if it is transmitted in sync with \(\hat{\bm{z}}_{t-\delta_{eu}}^{l}\). In addition, we assume that the agent's computation delay \(\delta_a\) and the control execution delay \(\delta_c\) are known constants, and the edge server continuously measures the downlink communication delay in real time. With this information, the agent remains aware of the \gls{e2e} delay \(\delta\), allowing it to dynamically adjust the prediction horizon.

\begin{figure*}[t]
    \begin{center}
    \includegraphics[width=0.95\linewidth]{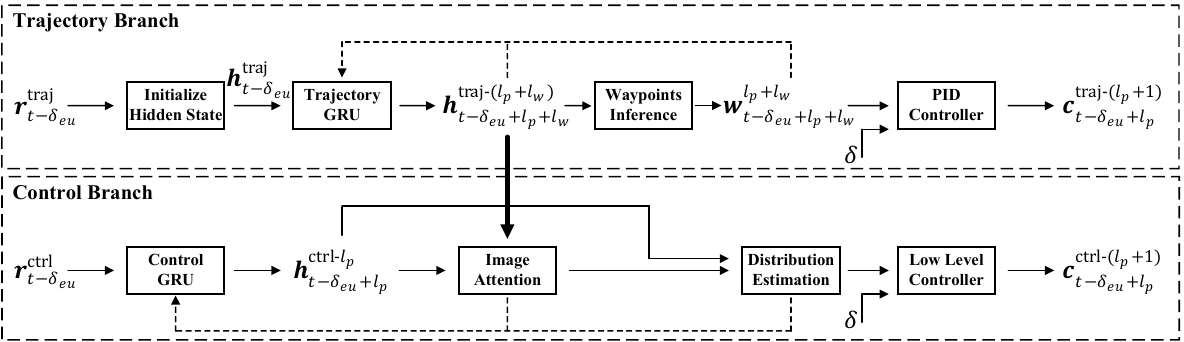}
    \end{center}
       \caption{The framework of the trajectory and control branch of DTCP.}
    \label{fig_ATCP_framework}
\end{figure*}

\subsection{Trajectory Branch}
The trajectory branch first generates planned waypoints, and then a low-level PID controller generates trajectory commands based on them. We define the function of the trajectory feature extractor as:
\begin{align}
    f_\text{feat-t}: \mathcal{Z}\times\mathcal{M}&\times\mathcal{H}\rightarrow\mathcal{R}_\text{traj} \notag\\
    &:(\hat{\bm{z}}_{t-\delta_{eu}}^l, \bm{m}_{t-\delta_{eu}}, \bm{h}_{t-\delta_{eu}})\mapsto \bm{r}_{t-\delta_{eu}}^\text{traj},
    \label{eq_traj_feat_map}
\end{align}
where \(\hat{\bm{z}}_{t-\delta_{eu}}^l, \bm{m}_{t-\delta_{eu}}\), and \(\bm{h}_{t-\delta_{eu}}\) represent the reconstructed \gls{jscc} symbols, state information, and channel state, respectively, corresponding to image \(\bm{x}_{t-\delta_{eu}}\) captured by the camera \(\delta_{eu}\) time slots ago.
At time \(t\), the trajectory feature on the edge server is denoted as \( \bm{r}_{t-\delta_{eu}}^{\text{traj}}\), as shown in \cref{fig_ATCP_framework}.
The trajectory hidden state \(\bm{h}_{t-\delta_{eu}}^{\text{traj}}\) of a \gls{gru} \cite{Cho_2014_LPR} is initialized with the trajectory feature. Then it auto-regressively generates the sequence of trajectory hidden states \(\bm{h}_{t-\delta_{eu}+l_p+l_w}^{\text{traj-}(l_p+l_w)} = (\bm{h}_{t-\delta_{eu}+l_p+l_w}^{\text{traj}}, \dots,\bm{h}_{t-\delta_{eu}+1}^{\text{traj}} ) \), where \(l_w\) denotes an extra prediction horizon for planned waypoints and \(l_p+l_w\) in the superscript represents the length of the sequence. Using a waypoint inference network, the planned \(l_p+l_w\) waypoints can be obtained from the sequence of trajectory hidden states, denoted as \(\bm{w}_{t-\delta_{eu}+l_p+l_w}^{l_p+l_w} = (\bm{w}_{t-\delta_{eu}+l_p+l_w}, \dots,\bm{w}_{t-\delta_{eu}+1})\). The initial waypoint \(\bm{w}_{t-\delta_{eu}}\) is defined as the origin. 

Each trajectory \(\bm{w}_{t-i+l_w}^{l_w+1} = (\bm{w}_{t-i+l_w}, \dots,\bm{w}_{t-i})\) with length \(l_w+1\) is processed by a PID controller to generate the predicted trajectory command \(\bm{c}^{\text{traj}}_{t-i}\), where \(i = \delta_{eu},\dots,\delta_{eu}-l_p\). Thus, the sequence of predicted trajectory commands branch is denoted as \( \bm{c}_{t-\delta_{eu}+l_p}^{\text{traj-}(l_p+1)} = (\bm{c}_{t-\delta_{eu}+l_p}^{\text{traj}}, \dots, \bm{c}_{t-\delta_{eu}}^{\text{traj}}) \). We defined the function of the trajectory branch as:
\begin{align}
    f_{\text{traj}}:\mathcal{R}_{\text{traj}}\rightarrow\mathcal{C}_{\text{cmd}}^{l_p+1}:\bm{r}_{t-\delta_{eu}}^{\text{traj}}\mapsto \bm{c}_{t-\delta_{eu}+l_p}^{\text{traj-}(l_p+1)}.
    \label{eq_traj_command_map}
\end{align}

\subsection{Control Branch}
As outlined in \cite{Wu_2022_TgC}, a control model that predicts current actions based solely on current inputs typically employs supervised training similar to behavior cloning, which assumes that the data is i.i.d. However, for autonomous driving, future states and commands are under the influence of
historical commands. To address this problem and deal with latency, the control branch predicts control commands in multiple steps in the future and obtains the desired commands based on the \gls{e2e} delay \(\delta\).

We defined the mapping of the reconstructed \gls{jscc} symbols to the control features as: 
\begin{align}
    f_\text{feat-c}: \mathcal{Z}\times\mathcal{M}&\times\mathcal{H}\rightarrow\mathcal{R}_\text{ctrl} \notag\\
    &:(\hat{\bm{z}}_{t-\delta_{eu}}^l, \bm{m}_{t-\delta_{eu}}, \bm{h}_{t-\delta_{eu}})\mapsto \bm{r}_{t-\delta_{eu}}^\text{ctrl}.
    \label{eq_ctrl_feat_map}
\end{align}
At time \(t\), the control hidden state \(\bm{h}_{t-\delta_{eu}}^{\text{ctrl}}\) is initialized with zero value and enters control \gls{gru} with the control feature \(\bm{r}_{t-\delta_{eu}}^{\text{ctrl}}\) to generate the next hidden state \(\bm{h}_{t-\delta_{eu}+1}^{\text{ctrl}}\). The hidden state of the control branch \( \bm{h}_{t-\delta_{eu}+1}^{\text{ctrl}} \) and the hidden state of the trajectory branch \( \bm{h}_{t-\delta_{eu}+1}^{\text{traj}} \) are used to estimate the important regions of the image by generating a binary mask that matches the shape of the image feature map from the middle layer of the control feature extractor \cite{Wu_2022_TgC}. This mask is then applied through element-wise multiplication with the feature map. The results of image attention are then used to generate the predicted control feature \(\bm{r}_{t-\delta_{eu}+1}^{\text{ctrl}}\) and the control command \( \bm{c}^{\text{ctrl}}_{t-\delta_{eu}+1} \). The next control \gls{gru} hidden state \(\bm{h}_{t-\delta_{eu}+2}^{\text{ctrl}}\) is obtained from the previous hidden state \(\bm{h}_{t-\delta_{eu}+1}^{\text{ctrl}}\) and the predicted control feature \(\bm{r}_{t-\delta_{eu}+1}^{\text{ctrl}}\). This process auto-regressively generates the sequence of control hidden states \( \bm{h}_{t-\delta_{eu}+l_p}^{\text{ctrl-}l_p} = (\bm{h}_{t-\delta_{eu}+l_p}^{\text{ctrl}}, \dots, \bm{h}_{t-\delta_{eu}+1}^{\text{ctrl}}) \), which is used to generate the sequence of predicted control features \( \bm{r}_{t-\delta_{eu}+l_p}^{\text{ctrl-}l_p} = (\bm{r}_{t-\delta_{eu}+l_p}^{\text{ctrl}}, \dots,\bm{r}_{t-\delta_{eu}+1}^{\text{ctrl}} ) \). Based on that, the sequence of predicted control commands \( \bm{c}_{t-\delta_{eu}+l_p}^{\text{ctrl-}(l_p+1)} = (\bm{c}_{t-\delta_{eu}+l_p}^{\text{ctrl}}, \dots, \bm{c}_{t-\delta_{eu}}^{\text{ctrl}}) \) is derived from a low-level controller, where \(\bm{c}_{t-\delta_{eu}}^{\text{ctrl}}\) is directly generated from the initial control feature \(\bm{r}_{t-\delta_{eu}}^{\text{ctrl}}\). The function of the trajectory branch is defined as:
\begin{align}
    f_{\text{ctrl}}:\mathcal{R}_{\text{ctrl}}\times\mathcal{H}_{\text{traj}}^{l_p}&\rightarrow\mathcal{C}_{\text{cmd}}^{l_p+1} \notag\\
    &:(\bm{r}_{t-\delta_{eu}}^{\text{ctrl}}, \bm{h}_{t-\delta_{eu}+l_p}^{\text{traj-}l_p})\mapsto \bm{c}_{t-\delta_{eu}+l_p}^{\text{ctrl-}(l_p+1)}.
    \label{eq_ctrl_command_map}
\end{align}

\subsection{Two Branch Combination}
To minimize the perceived \gls{e2e} delay \(\delta_r\), \(l_p \geq \delta\) must be satisfied, i.e., \( l_p - \delta_{eu} \geq \delta_{a}+\delta_{d}+\delta_{c}\). 
Because the trajectory branch and the control branch specialize in different driving scenarios, commands from the two branches are conditionally fused to obtain the combined command \( \bm{c}^{\text{comb}}_{t-\delta_{eu}+l_p} \).
This fusion depends on the driving situation -- whether the vehicle is turning or not. In addition, considering the trade-off between the prediction horizon and the reliability of the system, the predicted control is applied when the delay exceeds a certain threshold \(\delta_{T}\) for the turning situation. Otherwise, the robustness of the system can deal with the delay better than applying predicted commands. We define this combination function as:
\begin{align}
    f_\text{comb}: \mathcal{C}_\text{cmd}^{l_p+1}&\times\mathcal{C}_\text{cmd}^{l_p+1}\rightarrow\mathcal{C}_\text{cmd}\notag\\
    &:(\bm{c}_{t-\delta_{eu}+l_p}^{\text{traj-}(l_p+1)}, \bm{c}_{t-\delta_{eu}+l_p}^{\text{ctrl-}(l_p+1)})\mapsto \bm{c}_{t-\delta_{eu}+l_p}^{\text{comb}}.
    \label{eq_comb_map}
\end{align}
The combined command \(\bm{c}_{t-\delta_{eu}+l_p}^{\text{comb}}\) is denoted as:
\begin{align}
    \bm{c}_{t-\delta_{eu}+l_p}^{\text{comb}} = 
    \begin{cases}
      \lambda_{c}\cdot \bm{c}^{\text{traj}}_{t-\delta_{eu}+l_p}+ (1-\lambda_{c})\cdot \bm{c}^{\text{ctrl}}_{t-\delta_{eu}+l_p}, \\
      \qquad\qquad\qquad\qquad\quad\ \text{if turning and}~\delta\geq\delta_T,\\
      \lambda_{c}\cdot \bm{c}^{\text{traj}}_{t-\delta_{eu}}+ (1-\lambda_{c})\cdot \bm{c}^{\text{ctrl}}_{t-\delta_{eu}}, \\
      \qquad\qquad\qquad\qquad\quad\ \text{if turning and}~\delta<\delta_T,\\
      \lambda_{c}\cdot \bm{c}^{\text{ctrl}}_{t-\delta_{eu}}+ (1-\lambda_{c})\cdot \bm{c}^{\text{traj}}_{t-\delta_{eu}}, \ \quad \text{otherwise},
    \end{cases}
    \label{eq_command_combined}
\end{align}
where \( \lambda_{c} \in [0.5, 1]\) is a hyperparameter. The details of a complete cycle of communication, computing, and control of \gls{dtcp} and task-oriented \gls{jscc} are illustrated in \cref{alg_comb_command}.

\input{Algorithm/alg_comb_command}

\subsection{Loss function}
\label{subsec_loss_function}
We denote the estimated action corresponding to image \(\bm{x}_{t-\delta_{eu}}\) by 
\begin{align}
    \hat{\bm{a}}_{t-\delta_{eu}} = (v_{t-\delta_{eu}}, s_{t-\delta_{eu}}, &\bm{w}_{t-\delta_{eu}+l_p+l_w}^{l_p+l_w}, \bm{r}^{\text{traj}}_{t-\delta_{eu}}, \notag\\
    &\bm{c}_{t-\delta_{eu}+l_p}^{\text{ctrl-}{(l_p+1)}}, \bm{r}_{t-\delta_{eu}+l_p}^{\text{ctrl-}(l_p+1)}),
\end{align}
which consists of task-critical variables, where \( v_{t-\delta_{eu}} \) denotes the estimated target velocity and \( s_{t-\delta_{eu}} \) denotes the value of the extracted features. The corresponding ground-truth action is defined as: 
\begin{align}
    \bm{a}_{t-\delta_{eu}} = (\prescript{\text{ex}}{}{v}_{t-\delta_{eu}}, \prescript{\text{ex}}{}{s}_{t-\delta_{eu}}, 
    &\prescript{\text{ex}}{}{\bm{w}}_{t-\delta_{eu}+l_p+l_w}^{l_p+l_w}, 
    \prescript{\text{ex}}{}{\bm{r}}^{\text{traj}}_{t-\delta_{eu}}, \notag\\
    &\prescript{\text{ex}}{}{\bm{c}}_{t-\delta_{eu}+l_p}^{\text{ctrl-}{(l_p+1)}}, \prescript{\text{ex}}{}{\bm{r}}_{t-\delta_{eu}+l_p}^{\text{ctrl-}(l_p+1)}), 
\end{align}
which is collected from expert agents or human drivers.

The loss function of the trajectory branch is defined as follows:
\begin{align}
    \mathcal{L}_{\text{traj}} = \|\bm{w}_{t-\delta_{eu}+l_p+l_w}^{l_p+l_w} - &\prescript{\text{ex}}{}{\bm{w}}_{t-\delta_{eu}+l_p+l_w}^{l_p+l_w}\|_1 \notag\\
    &+ \lambda_{\text{feat}}\| \bm{r}^{\text{traj}}_{t-\delta_{eu}} - \prescript{\text{ex}}{}{\bm{r}}^{\text{traj}}_{t-\delta_{eu}}\|_2,
    \label{eq_loss_traj}
\end{align}
where \( \lambda_{\text{feat}} \) is a hyperparameter, $\|\cdot\|_{1}$ denotes the $\ell_{1}$-norm, $\|\cdot\|_{2}$ denotes the Euclidean distance ($\ell_{2}$-norm).

For the control branch, the distribution of the control action is modeled as a beta distribution \cite{Wu_2022_TgC}. The loss function of the control branch is defined as follows:
\begin{align}
    \mathcal{L}_{\text{ctrl}}
    =\frac{1}{l_p+1}&\sum_{i=t-\delta_{eu}}^{t-\delta_{eu}+l_p} D_{\text{KL}}(\mathcal{B}e(\bm{c}_{i}^{\text{ctrl}})\|\mathcal{B}e(\prescript{\text{ex}}{}{\bm{c}}_{i}^{\text{ctrl}}))  \notag\\
    & + \lambda_{\text{feat}}\|\bm{r}_{t-\delta_{eu}+l_p}^{\text{ctrl-}(l_p+1)} - \prescript{\text{ex}}{}{\bm{r}}_{t-\delta_{eu}+l_p}^{\text{ctrl-}(l_p+1)}\|_{2},
    \label{eq_loss_ctrl}
\end{align}
where $\mathcal{B}e(\cdot)$ denotes the beta distribution. Furthermore, an auxiliary function is used to measure the accuracy of the estimated current speed and value that is obtained from the speed head and the value head, respectively, to help the agent make decisions \cite{Wu_2022_TgC}. The auxiliary function is defined as:
\begin{align}
    \mathcal{L}_{\text{aux}} = \lambda_{\text{value}}\|v_{t-\delta_{eu}} &- \prescript{\text{ex}}{}{v}_{t-\delta_{eu}}\|_{1} \notag\\
    &+ \lambda_{\text{speed}}\|s_{t-\delta_{eu}}-\prescript{\text{ex}}{}{s}_{t-\delta_{eu}}\|_{2},
    \label{eq_loss_aux}
\end{align}
where \(\lambda_{\text{value}}\) and \(\lambda_{\text{speed}}\) are hyperparameters. Thus, the overall loss function of the \gls{dtcp} is defined as:
\begin{align}
    \mathcal{L}_{\text{DTCP}} = \lambda_{\text{traj}}\mathcal{L}_{\text{traj}} + \lambda_{\text{ctrl}}\mathcal{L}_{\text{ctrl}} + \lambda_{\text{aux}}\mathcal{L}_{\text{aux}},
    \label{eq_loss_DTCP}
\end{align}
where $\lambda_{\text{traj}}$, $\lambda_{\text{ctrl}}$, and $\lambda_{\text{aux}}$ are hyperparameters. The design of the loss functions in \cref{eq_loss_traj}, \cref{eq_loss_ctrl}, and \cref{eq_loss_aux} follows a consistent principle: combining an output loss and a feature loss through a weighted summation. We consider these weights essential because the two types of variables (e.g., waypoints and trajectory features) typically have different scales and ranges, requiring proper balancing to ensure meaningful contributions from each term. For the overall loss function in \cref{eq_loss_DTCP}, the weights ($\lambda_{\text{traj}}$, $\lambda_{\text{ctrl}}$, $\lambda_{\text{aux}}$) are carefully chosen to ensure that each component contributes appropriately to the task objective, aligning with the goal of achieving better system performance.

\subsection{Joint Training}
To jointly train the \gls{dtcp} and task-oriented \gls{jscc}, we employ imitation learning, specifically through behavior cloning. In this approach, the agent learns to perform tasks by replicating the actions of experts based on a dataset of expert demonstrations. Behavior cloning works by directly mapping observed states to corresponding actions, allowing the agent to learn a policy that mirrors the expert’s behavior. This approach is particularly effective in scenarios where a large amount of labeled data is available, allowing the agent to generalize from the expert's actions to similar situations encountered during autonomous driving.

Assuming the posterior $q_{\psi}(\bm{a}|\hat{\bm{z}}, \bm{m}, \bm{h})$ follows a Gaussian distribution $\mathcal{N}(\mu_{\psi}(\hat{\bm{z}}, \bm{m}, \bm{h}), \sigma_{\text{const}}^{2}I)$, where $\mu_{\psi}(\hat{\bm{z}}, \bm{m}, \bm{h})$ maps reconstructed \gls{jscc} symbols $\hat{\bm{z}}$, state information \(\bm{m}\), and channel state \(\bm{h}\) to the mean of a Gaussian distribution and $\sigma_{\text{const}}$ is a constant, we can derive the following expression: 
\begin{align}
    -\log{q_{\psi}(\bm{a}|\hat{\bm{z}}, \bm{m}, \bm{h})} \sim\frac{1}{2\sigma_{\text{const}}^{2}}\|\bm{a}-\mu_{\psi}(\hat{\bm{z}}, \bm{m}, \bm{h}) \|^2_2,
    \label{eq_log2normal}
\end{align}
where $\mu_{\psi}(\hat{\bm{z}}, \bm{m}, \bm{h})=\hat{\bm{a}}$.
\cref{eq_log2normal} shows that $-\log q_{\psi}(\bm{a}|\hat{\bm{z}}, \bm{m}, \bm{h})$ can serve as a distance metric, analogous to the square of the $\ell^2$-norm. From this perspective, we heuristically regard the loss function of \gls{dtcp} as an extension of the first term in \cref{eq_VIB_realization}, thus we can jointly optimize \gls{dtcp} with task-oriented communication as follows:
\begin{align}
    \mathcal{L}_{\text{VIB}}':= \frac{1}{K_{b}}\sum_{i=1}^{K_{b}} 
\bigl\{&
\mathcal{L}_\text{DTCP} + \beta D_{\text{KL}}(p_{\phi}(\hat{\bm{z}} | \bm{x}_{i}) \| q(\hat{\bm{z}}))
\bigr\}.
\label{eq_joint_loss}
\end{align}
The joint training process of proposed task-oriented co-design is shown in \cref{alg_joint_train}. 

\input{Algorithm/alg_joint_train}

%% file: Algorithm/alg_comb_command.tex
\begin{algorithm}[t]
\caption{Communication, Computing, and Control of DTCP and Task-Oriented JSCC.}
\label{alg_comb_command}
\begin{algorithmic}[1]
\State \textbf{Initialization}: Load the pre-trained parameter \(\phi\) for JSCC encoder (\(f_e\)) and parameter \(\psi\) for DTCP (\(f_\text{feat-t}, f_\text{feat-c}, f_\text{traj}\), and \(f_\text{ctrl}\)). 

\State \textbf{Vehicle}:
\State \quad\ \ At time \(t-\delta_{eu}\), capture image \(\bm{x}_{t-\delta_{eu}}\) and generate 
\Statex \quad\ \ state information \(\bm{m}_{t-\delta_{eu}}\).
\State \quad\ \ At time \(t-\delta_u\), generate selected JSCC symbols:
\Statex \quad\ \ \(\bm{z}_{t-\delta_{eu}}^{l} \leftarrow f_e(\bm{x}_{t-\delta_{eu}})\).

\State \textbf{Edge Server}:
\State \quad\ \ At time \(t\), receive reconstructed JSCC symbols
\Statex \quad\ \ \(\hat{\bm{z}}_{t-\delta_{eu}}^{l}\leftarrow f_h(\bm{z}_{t-\delta_{eu}}^{l})\) and state information \(\bm{m}_{t-\delta_{eu}}\). 
\Statex \quad\ \ Measure corresponding channel state \(\bm{h}_{t-\delta_{eu}}\).
\State \quad\ \ Generate trajectory feature: 
\Statex \quad\ \ \(\bm{r}_{t-\delta_{eu}}^{\text{traj}} \leftarrow f_\text{feat-t}(\hat{\bm{z}}_{t-\delta_{eu}}^l, \bm{m}_{t-\delta_{eu}}, \bm{h}_{t-\delta_{eu}})\).
\State \quad\ \ Generate sequence of trajectory command:
\Statex \quad\ \ \(\bm{c}_{t-\delta_{eu}+l_p}^{\text{traj-}(l_p+1)}\leftarrow f_\text{traj}(\bm{r}_{t-\delta_{eu}}^{\text{traj}})\), and sequence of hidden 
\Statex \quad\ \ state \(\bm{h}_{t-\delta_{eu}+l_p}^{\text{traj-}l_p}\).
\State \quad\ \ Generate control feature:
\Statex \quad\ \ \(\bm{r}_{t-\delta_{eu}}^{\text{ctrl}} \leftarrow f_\text{feat-c}(\hat{\bm{z}}_{t-\delta_{eu}}^l, \bm{m}_{t-\delta_{eu}}, \bm{h}_{t-\delta_{eu}})\).
\State \quad\ \ Generate sequence of control command:
\Statex \quad\ \ \(\bm{c}_{t-\delta_{eu}+l_p}^{\text{ctrl-}(l_p+1)}\leftarrow f_\text{ctrl}(\bm{r}_{t-\delta_{eu}}^{\text{ctrl}}, \bm{h}_{t-\delta_{eu}+l_p}^{\text{traj-}l_p})\).
\State \quad\ \ At time \(t+\delta_a\), generate combined command:
\Statex \quad\ \ \(\bm{c}_{t-\delta_{eu}+l_p}^{\text{comb}}\leftarrow f_\text{comb}(\bm{c}_{t-\delta_{eu}+l_p}^{\text{traj-}(l_p+1)}, \bm{c}_{t-\delta_{eu}+l_p}^{\text{ctrl-}(l_p+1)})\).
\State \textbf{Vehicle}:
\State \quad\ \ At time \(t+\delta_a+\delta_d\), receive command \(\bm{c}_{t-\delta_{eu}+l_p}^{\text{comb}}\).
\State \quad\ \ At time \(t+\delta_{a}+\delta_{d}+\delta_{c}\), vehicle is controlled by the
\Statex \quad\ \ command \(\bm{c}_{t-\delta_{eu}+l_p}^{\text{comb}}\). 

\end{algorithmic}
\end{algorithm}

%% file: Algorithm/alg_joint_train.tex
\begin{algorithm}[t]
\caption{Joint Training of DTCP and Task-Oriented JSCC.}
\label{alg_joint_train}
\begin{algorithmic}[1]
\Statex \textbf{Initialization}: Initialize the neural network parameters $\phi$ and $\psi$.

\State \textbf{Input}: Image dataset $\mathcal{X}$ with corresponding ground-truth agent output $\mathcal{A}$ and state information $\mathcal{M}$.

\While{not converged}
    \State Sample mini-batch $\{(\bm{a}_i, \bm{x}_i)\}_{i=1}^{K_b}$ from $\mathcal{A}$ and $\mathcal{X}$ with 
    \Statex \quad \, corresponding state information $\{\bm{m}_i\}_{i=1}^{K_b}$ from $\mathcal{M}$.
    
    \For{sample $i=1,\dots,K_b$}
        \State Encode image to JSCC symbols:
        \(\bm{z}_{i}\leftarrow f_{e}(\bm{x_i})\).
        \State Transmit JSCC symbols through channel and apply 
        \Statex \qquad \ \ \ equalization: \(\hat{\bm{z}}_i\leftarrow f_{h}(\bm{z}_i)\).
        \State Estimate mean and standard deviation: 
        \Statex \qquad \ \ \ \(\bm{\mu}_i\leftarrow f_{\bm{\mu}}(\hat{\bm{z}}_i)\), \(\bm{\sigma}_i\leftarrow f_{\bm{\sigma}}(\hat{\bm{z}}_i)\).
        \State Compute KL divergence \(D_{\text{KL}}(p_{\phi}(\hat{\bm{z}} | \bm{x}_{i}) \| q(\hat{\bm{z}}))\).
        \State Generate estimated action: \(\hat{\bm{a}}_i\leftarrow f_{a}(\hat{\bm{z}}_i, \bm{m}_i, \bm{h}_i)\).
        \State Compute DTCP loss based on \cref{eq_loss_DTCP}.
    \EndFor
    
    \State Compute joint loss \(\mathcal{L}_\text{VIB}'\) of this mini-batch based on 
    \Statex \quad \, \cref{eq_joint_loss}. 
    \State Update neural network parameters:
    \(\phi \overset{+}\leftarrow -\nabla_{\phi}\mathcal{L}_\text{VIB}' \),
    \Statex \quad \, \(\psi \overset{+}\leftarrow -\nabla_{\psi}\mathcal{L}_\text{VIB}' \).
\EndWhile

\end{algorithmic}
\end{algorithm}

%% file: 3_experiment.tex
\section{Performance Evaluation}
\label{sec_result}
In this section, we present a case study of our proposed task-oriented co-design framework. The evaluation is carried out within the simulator \gls{carla}, which offers a variety of urban environments that closely mimic real-world traffic scenarios.

\subsection{Experimental Setup}
\label{subsec_exp_setup}
\paragraph{Dataset}
We utilize the well-structured dataset provided by \cite{Wu_2022_TgC}, which consists of images (height = 256, width = 900, channels = 3) captured from various urban environments, along with the corresponding vehicle state information. Specifically, the dataset contains \(K_a = 189524\) images from four maps (Town01, Town03, Town04, and Town06) for training, and 27201 images from four different maps (Town02, Town05, Town07, and Town10) for testing. This well-structured dataset allows us to effectively train and validate the proposed framework across a range of real-world-like scenarios.

To train \gls{dtcp} using behavior cloning, we use Roach \cite{Zhang_2021_EtE} as the expert agent in our experiments. Roach is a highly capable autonomous driving agent that relies on \gls{bev} as input. Since \gls{bev} data are challenging to collect in real time for real-world autonomous driving, this highlights the importance of training autonomous driving agents using data from standard sensors. Our proposed \gls{dtcp}, equipped with only one camera, demonstrates strong potential for practical deployment in real-world scenarios.

\paragraph{Evaluation} 
The experiment is designed to evaluate the driving performance of the proposed task-oriented co-design framework against established baselines under varying communication conditions.
These conditions include significant communication latency, constrained bandwidth, and the presence of noisy fading channels. The baselines for comparison include three widely recognized image coding techniques: 1) JPEG \cite{Wallace_1992_TJs}; 2) JPEG2000 \cite{Taubman_2002_JIc}; and 3) BPG \cite{BPG}. Each coding method is followed by (2048, 6144) \gls{ldpc} codes with a 64-QAM digital modulation scheme.

In addition, two \gls{jscc}-based methods, referred to as ``JSCC-AE'' \cite{Bourtsoulatze_2019_DJS} and ``JSCC-VAE'' \cite{Saidutta_2021_JSC}, are also included as baselines. JSCC-AE is a seminal work that introduced the concept of joint source-channel coding without relying on explicit separate codes for compression or error correction, making it a foundational approach in this area. Based on this, JSCC-VAE offers robustness against variations in channel conditions, further enhancing its practical applicability. These methods focus on accurately reconstructing the image at the edge server, but do not co-design with the autonomous driving agent (\gls{dtcp}). Furthermore, the baseline includes \cite{Wu_2022_TgC}, which performs driving tasks using uncompressed images, denoted as ``TGCP.''

Driving performance is quantified using the established driving score metric\footnote{\url{https://leaderboard.carla.org/}} of CARLA, which evaluates the vehicle's ability to follow predefined waypoints, reach target destinations, and comply with traffic regulations. To ensure robustness, each experiment is repeated three times on a selected route in Town05, under four distinct weather conditions: clear noon, cloudy sunset, soft rain at dawn, and heavy rain at night. For a more intuitive comparison, we select road sections where TGCP achieves a perfect driving score of 100.

\paragraph{Parameters Settings}
For the task-oriented \gls{jscc} encoder, we configure the dimension of the \gls{jscc} symbols to \( l_z = 1024 \), achieving a significant low bandwidth compression ratio of \(l_z/l_x \approx 0.0015\). The average power constraint \(P_\text{target}\) for \gls{jscc} symbols is fixed at 1. The predefined Gaussian distribution is assumed to be $q(\hat{\bm{z}}) \sim \mathcal{N}(0,I)$.
In addition, ``JSCC-AE'' and ``JSCC-VAE'' use the same network structure as the proposed task-oriented JSCC for fair comparisons. 

\begin{figure*}[t]
    \begin{center}
    \includegraphics[width=0.49\linewidth, angle=90]{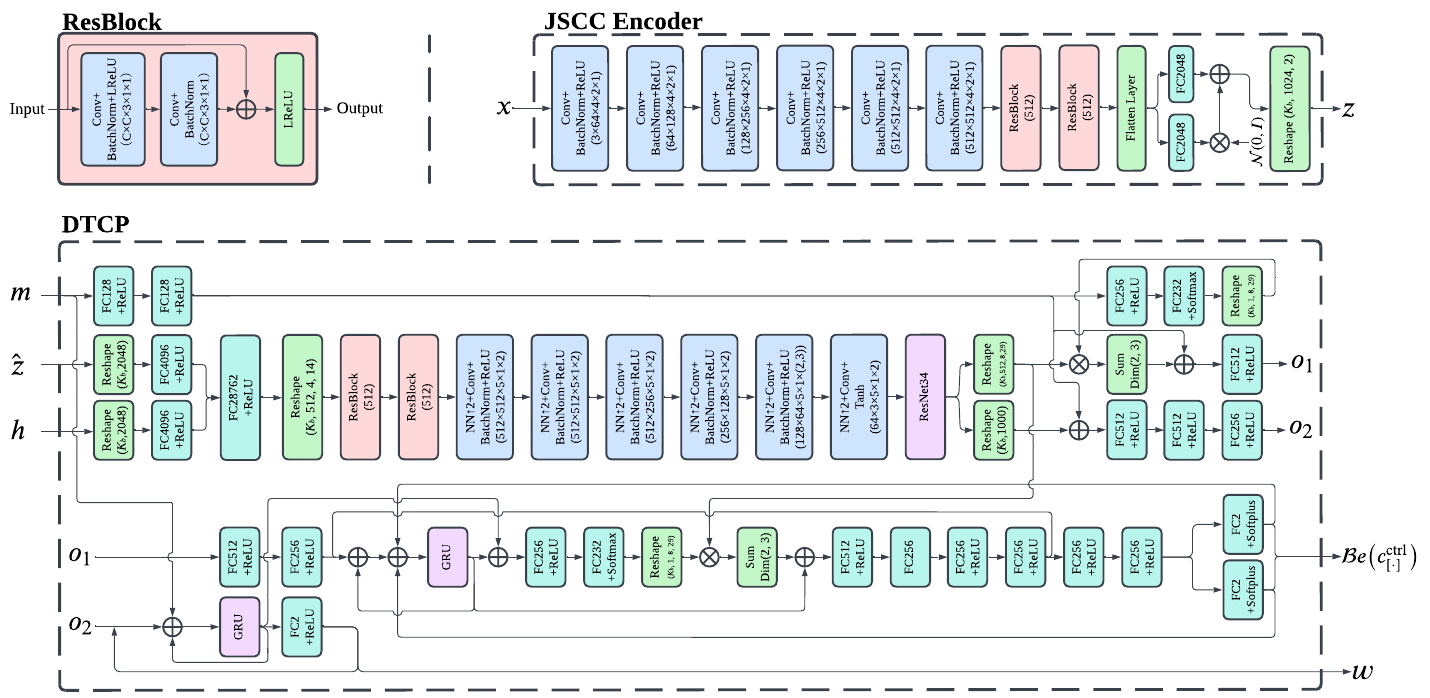}
    \end{center}
       \caption{Neural network architecture of the proposed JSCC encoder and DTCP. The main components are annotated as follows: \textbf{Conv}: Convolutional layer, with parameters specified as (\textit{input channel size} × \textit{output channel size} × \textit{kernel size} × \textit{stride} × \textit{padding}). \textbf{FC}: Fully-connected layer, where the following number indicates the output dimensions. \textbf{NN$\uparrow$2}: Nearest neighbor upsampling. \textbf{ResBlock}: Residual block, with parameters specifying the input and output channel sizes. \textbf{Reshape}: Reshaping layer, with parameters specifying the target dimensions. \textbf{LReLU}: Leaky ReLU activation function with $\alpha = 0.2$. \textbf{Softplus}: Softplus activation function. \textbf{Sum Dim(2,3)}: Summation operation performed along dimensions 2 and 3 \cite{Wu_2022_TgC}. \textbf{GRU}: Gated Recurrent Unit (GRU) \cite{Cho_2014_LPR}. Connection points $o_1$ and $o_2$ represent linked points, specifically, all instances of $o_1$ are interconnected, as are all instances of $o_2$.}
    \label{fig_DNN_structure}
\end{figure*}

For \gls{dtcp}, the parameters are configured as follows: \(\lambda_{c} = 0.7\), \(\lambda_{\text{feat}} = 0.05\), \(\lambda_{\text{value}} = 0.001\), \(\lambda_{\text{speed}} = 0.05\), and \(\lambda_{\text{traj}} = \lambda_{\text{ctrl}} = \lambda_{\text{aux}} = 1\). The values of $\lambda_{[\cdot]}$  were selected based on our preliminary tests, which ensure a stable training process and achieve relatively optimal driving performance. In our preliminary tests, we observed that excessively large ($\beta > 0.01$) or small ($\beta < 0.000001$) values of $\beta$ can disrupt the balance between the IB terms, leading to training instability and crashes. To address this, we set $\beta=0.0001$ for jointly training \gls{jscc} encoder and \gls{dtcp} under the IB objective. This $\beta$ value creates a reasonable balance between the two IB terms, ensuring stable training and achieving good overall performance. For the mini-batch, we set \(K_b=32\). Moreover, the duration of each time slot \(\tau\) is synchronized with the simulation time step of \gls{carla}, which is 0.05 seconds. The neural network architectures of the proposed \gls{jscc} encoder and \gls{dtcp} are shown in \cref{fig_DNN_structure}.

For the \gls{ofdm} system, the parameters are set to: $N_{\text{sub}}=12$, $N_\text{path}=8$, $\gamma=4$, and the length of the cyclic prefix (CP) is 3.

\subsection{Evaluation on CARLA}
\label{subsec_eva_CARLA}
\subsubsection{Constrained Bandwidth Compression Ratio}
\label{subsubsec_bandwidth}
\begin{figure}[t]
    \begin{center}
    \includegraphics[width=0.95\linewidth]{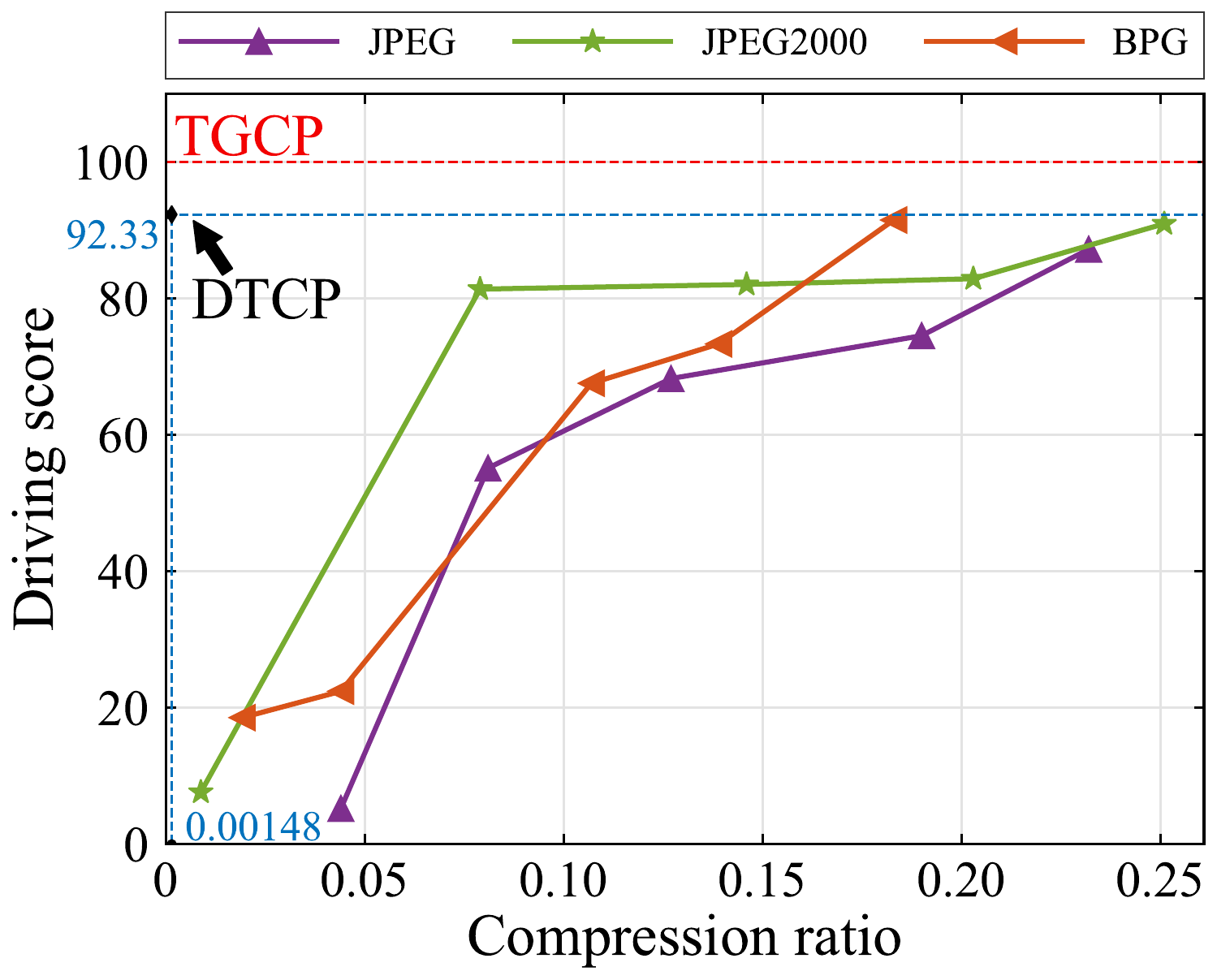}
    \end{center}
       \caption{Driving scores of traditional coding methods with varied bandwidth compression ratios under OFDM channel with SNR = 20 dB.}
    \label{fig_bandComp_ratio}
\end{figure}

The effect of the bandwidth compression ratio on the driving score is illustrated in \cref{fig_bandComp_ratio}. When the required driving score is 90, the proposed \gls{dtcp} secures substantial reductions in bandwidth usage by at least 99.19\% compared to traditional coding methods. In contrast, if bandwidth compression ratios are drastically reduced for traditional coding methods (i.e., less than 0.05), the corresponding reduction in image quality leads to a severe degradation in driving performance, with driving scores struggling to exceed 20.
This comparison shows the limits of conventional approaches under extreme bandwidth constraints and showcases the superior adaptability of our task-oriented co-design framework in such challenging scenarios.

\subsubsection{Noisy Fading Channel}
\label{subsubsec_Rayleigh}
\begin{figure}[t]
    \begin{center}
    \includegraphics[width=0.95\linewidth]{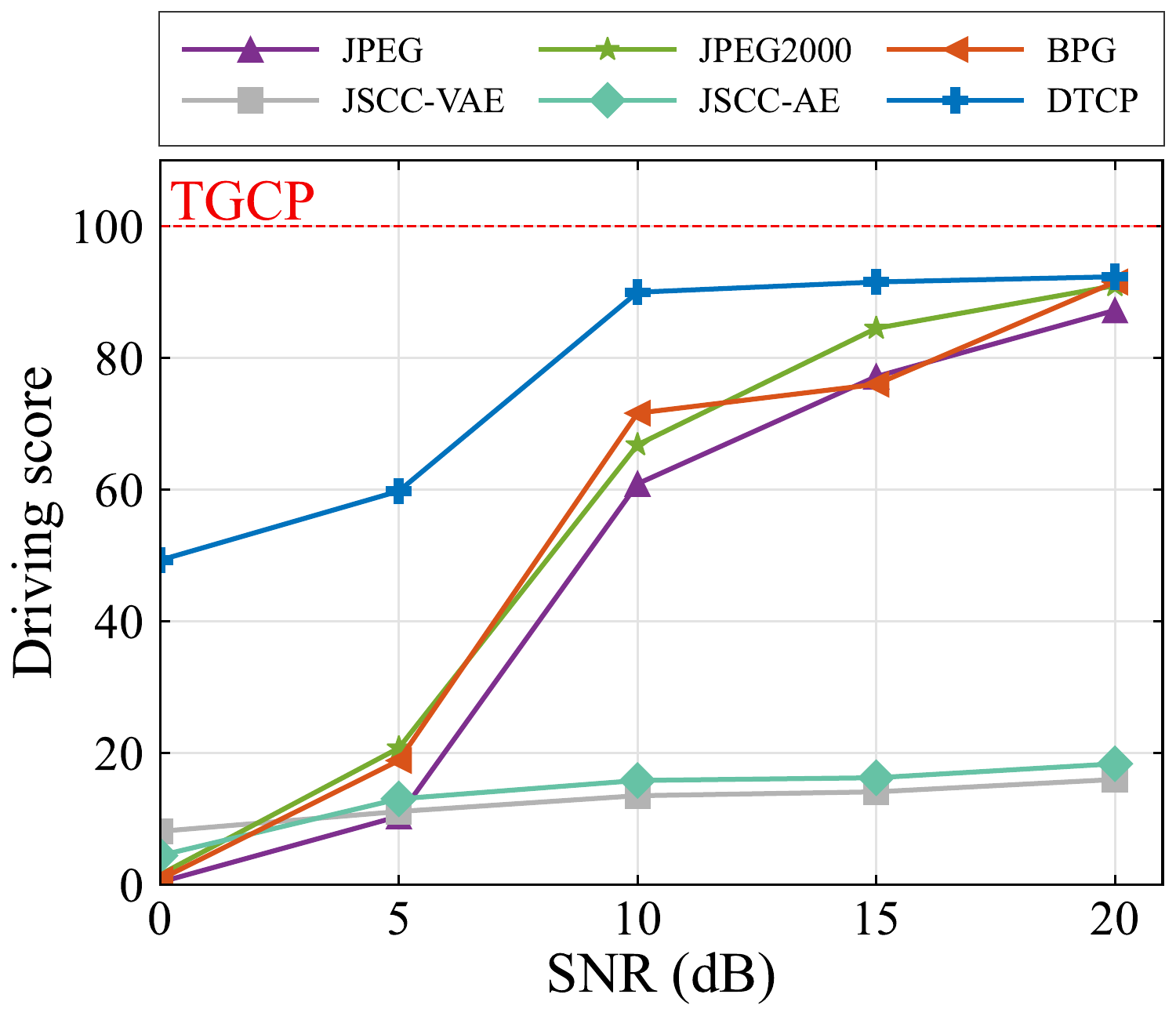}
    \end{center}
       \caption{Driving scores with varied SNRs under OFDM channel.}
    \label{fig_score_comp}
\end{figure}

In \cref{fig_score_comp}, we analyze variations in driving performance as a function of \gls{snr} under an \gls{ofdm} channel. Drawing from the findings in the constrained bandwidth compression ratio experiment, we set the bandwidth compression ratios to 0.232 for JPEG, 0.251 for JPEG2000, and 0.183 for BPG, where traditional methods perform comparably to \gls{dtcp}, with driving scores consistently higher than 90, ensuring a fair comparison.

When $\text{SNR}\geq 15 \ \text{dB}$, the proposed \gls{dtcp} framework performs similarly to traditional coding methods. When $\text{SNR} = 10 \ \text{dB}$, JPEG, JPEG2000, and BPG occasionally encounter decoding errors, leading to driving scores below 72, while \gls{dtcp} still maintains a driving score above 89.
As \gls{snr} drops below 5 dB, the \gls{dtcp} framework continues to maintain robust driving performance, with scores remaining above 49.
Specifically, the \gls{dtcp} framework achieves driving scores of 59.78 at \gls{snr} = 5 dB and 49.29 at \gls{snr} = 0 dB. In contrast, severe noise significantly hampers the performance of systems utilizing traditional coding methods when \gls{snr} is lower than 5 dB, causing frequent decoding failures and dramatically low driving scores (below 21 when $\text{SNR} = 5 \ \text{dB}$ and below 2 when $\text{SNR} = 0 \ \text{dB}$). Additionally, both the JSCC-AE and JSCC-VAE methods consistently produce driving scores below 20 across all \gls{snr} levels, highlighting the importance of task-oriented co-design in transmitting task-critical information.

These results underscore the resilience of our task-oriented co-design framework under adverse noise conditions, demonstrating its ability to maintain effective performance even in highly challenging environments. Moreover, the framework achieves excellent scores under regular channel conditions while simultaneously achieving bandwidth savings of at least 99.19\%. This highlights the efficiency and robustness of the \gls{dtcp} approach, making it a viable solution for real-world scenarios where communication channels are unreliable and constrained.

\subsubsection{Selection of JSCC symbols}
Given the characteristics of \gls{jscc}, symbols with relatively low energy are particularly vulnerable to noise. To optimize the trade-off between bandwidth compression and driving performance, we explore the selection of generated \gls{jscc} symbols, aiming to further reduce the bandwidth while maintaining the required driving performance.

\begin{figure}[t]
    \begin{center}
    \includegraphics[width=0.95\linewidth]{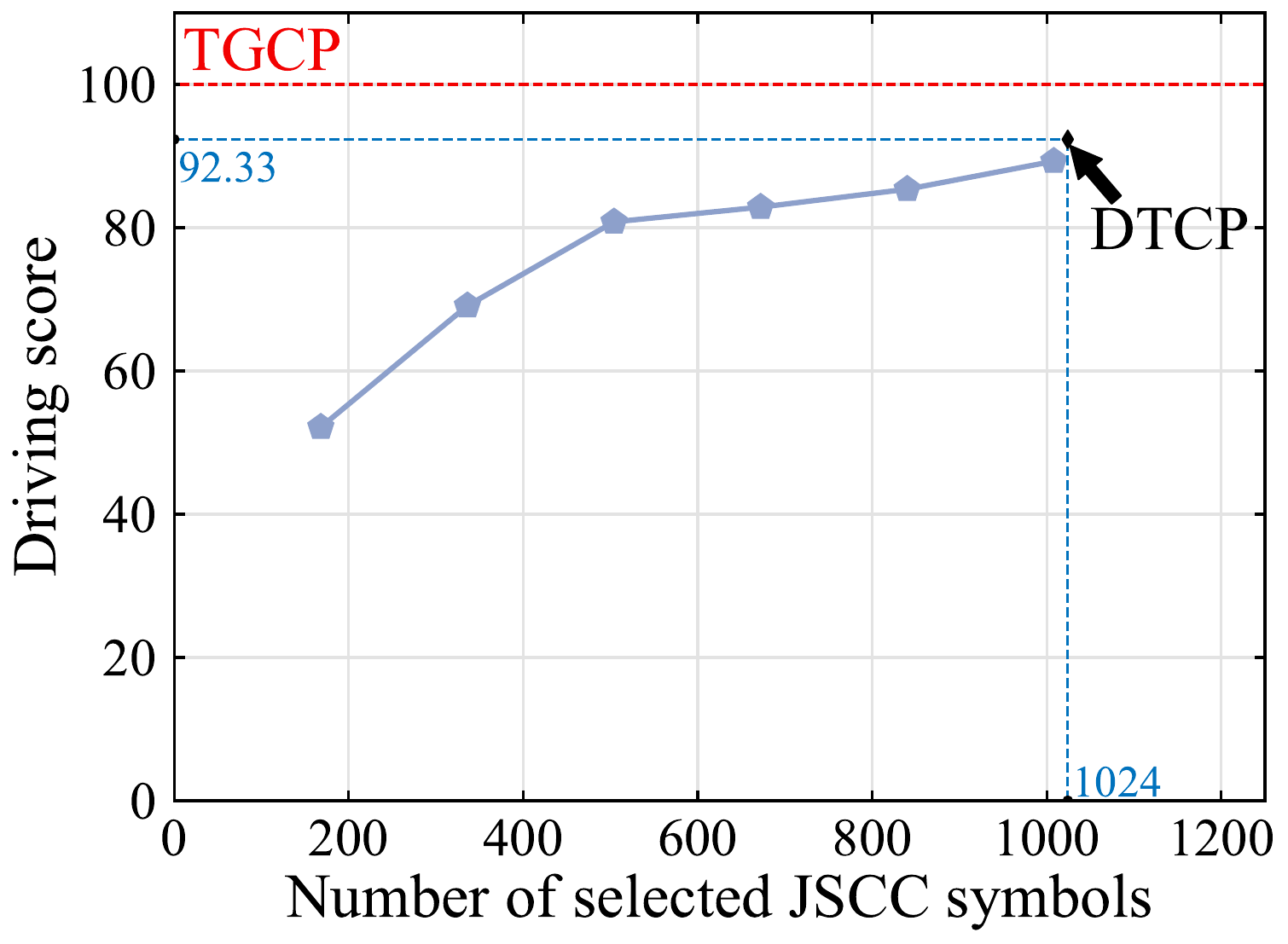}
    \end{center}
       \caption{Driving scores with varied selected JSCC symbols under OFDM channel with SNR = 20 dB.}
    \label{fig_z_selected_comp}
\end{figure}

As shown in \cref{fig_z_selected_comp}, the number of selected \gls{jscc} symbols varies from 168 to 1008, in increments of 168, while the corresponding bandwidth compression ratio varies from 0.00024 to 0.00146.
The choice of 168 as the incremental step size is based on the structure of a 5G resource block, which consists of 12 subcarriers and 14 OFDM symbols per slot, totaling 168 resource elements \cite{GPP_2020_5NP}. This value is a natural fit for our simulation setup, as it aligns with the granularity of resource allocation in modern cellular networks, making the results more relevant for real-world applications.

The driving score exhibits a gradual decline (from 89.28 to 80.81) as the number of selected \gls{jscc} symbols decreases from 1008 to 504. However, the driving score drops sharply (from 80.81 to 52.15) when the number of selected \gls{jscc} symbols is reduced further from 504 to 168. This significant drop suggests that high-energy \gls{jscc} symbols are more critical to task performance, as they carry essential information required for accurate decision-making in autonomous driving tasks.

Our proposed \gls{dtcp} framework demonstrates the ability to maintain a driving score above 80 by transmitting only the top 504 high-energy \gls{jscc} symbols. This selective transmission strategy demonstrates the potential to reduce communication overhead by 50.78\% compared to transmitting 1024 \gls{jscc} symbols. Achieving this reduction depends on adequately mitigating index transmission overhead, which could be addressed by applying techniques such as VL-VFE \cite{Shao_2022_LTO}. This approach not only optimizes bandwidth usage but also ensures robust driving performance.

\subsubsection{Compensate Perceived E2E Delay}
\begin{figure}[t]
    \begin{center}
    \includegraphics[width=0.95\linewidth]{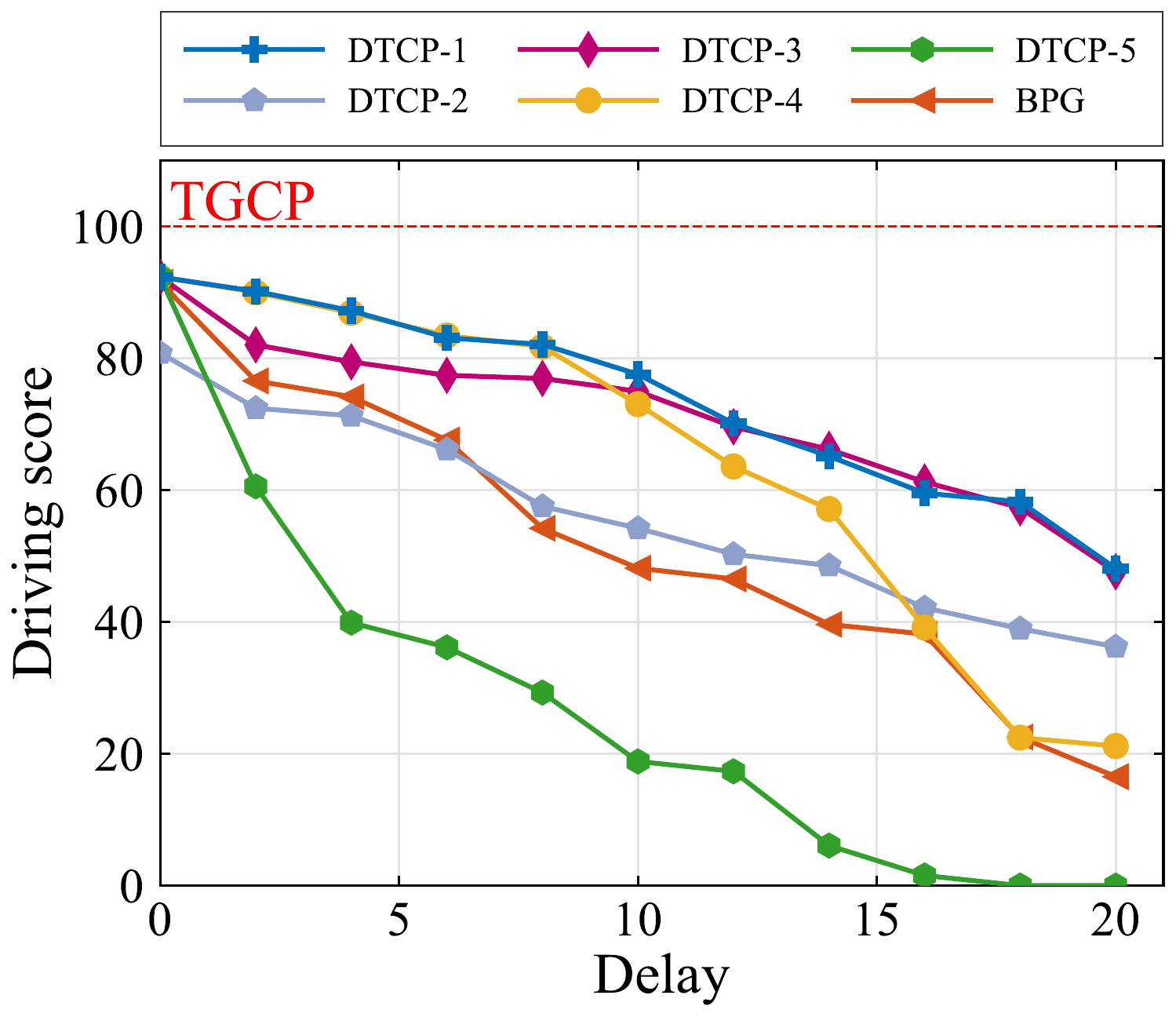}
    \end{center}
       \caption{Driving scores with varied delays under OFDM channel with SNR = 20 dB.}
    \label{fig_delay_comp}
\end{figure}

The impact of communication delays on driving performance using the \gls{dtcp} framework is presented in \cref{fig_delay_comp}. The delay ranges from 0 to 20 time slots, increasing by increments of 2 time slots. We evaluate five distinct configurations within the \gls{dtcp} framework:
\begin{itemize}
    \item \textbf{DTCP-1}: Transmits all \gls{jscc} symbols and generates commands based on \cref{eq_command_combined} with parameters \(l=1024\), \(l_p=\delta\), and \(\delta_T=10\). This is also the default configuration of \gls{dtcp} in previous experiments.
    \item \textbf{DTCP-2}: Selects 504 \gls{jscc} symbols for transmission, generating commands according to \cref{eq_command_combined} with \(l=504\), \(l_p=\delta\), and \(\delta_T=10\).
    \item \textbf{DTCP-3}: Transmits all \gls{jscc} symbols and always generates predicted commands for the turning situation (\(l=1024\), \(l_p=\delta\), and \(\delta_T=0\)).
    \item \textbf{DTCP-4}: Transmits all \gls{jscc} symbols but generates commands without prediction (\(l=1024\), \(l_p=0\), and \(\delta_{T} \rightarrow \infty\)).
    \item \textbf{DTCP-5}: Transmits all \gls{jscc} symbols and always generates predicted commands for all situations (\(l=1024\), \(l_p=\delta\), and \(\delta_T=0\)).
\end{itemize}
In this experiment, BPG with a bandwidth compression ratio of 0.183 is used as a representative baseline.

The results show that DTCP-5 experiences a steep decline in driving scores, falling below 61 even with a delay of just 2 time slots, and continues to decrease with increasing delay. In addition, DTCP-5 also performs worse than BPG in the presence of delays. These results indicate that relying exclusively on predicted commands is unreliable, particularly in non-turning scenarios.

When the delay is less than 10 time slots, DTCP-4 manages to maintain a driving score greater than 80 without relying on predicted commands. However, beyond 10 time slots, driving performance sharply declines, highlighting the limitations of unpredicted commands in high-latency conditions. In particular, when the delay is less than 8 time slots, the superior performance of DTCP-4 compared to DTCP-3 demonstrates that receiving accurate commands, even with some latency, is more critical than receiving inaccurate predicted commands under low latency. In contrast, when the delay exceeds 10 time slots, DTCP-3 outperforms DTCP-4, showing that adopting predicted commands becomes more effective in high-latency environments.

DTCP-1 combines the advantages of DTCP-3 and DTCP-4, offering the most balanced performance by dynamically switching between unpredicted and predicted commands based on \gls{e2e} delay. It outperforms DTCP-4 and BPG significantly by 20.39 and 21.38 points at \(\delta=16\), 35.78 and 35.69 points at \(\delta=18\), and 26.95 and 31.59 points at \(\delta=20\), respectively. Furthermore, DTCP-1 outperforms DTCP-3 when the delay is less than 10 time slots.

DTCP-2, while leading to an average reduction of 17 points compared to DTCP-1, shows the potential to preserve 50.78\% of communication resources. Despite the decrease in driving performance, DTCP-2 still maintains a driving score above 50, even with delays of up to 12 time slots. Furthermore, the driving scores of DTCP-2 exceed BPG and DTCP-4 when the delay is greater than 8 and 16 time slots, respectively, highlighting the efficiency of DTCP-2 in managing significant delays and offering a viable trade-off between communication bandwidth and driving performance.

%% file: 4_conclusion.tex
\section{Conclusion}
\label{sec_conclusion}
In this paper, we introduced a novel task-oriented co-design framework that integrates communication, computing, and control, specifically tailored for edge-enabled industrial \gls{cps}. By leveraging task-oriented \gls{jscc} through \gls{vib} theory, our approach effectively discards task-agnostic information, resulting in significant savings in communication bandwidth. Furthermore, with the incorporation of delay awareness into the trajectory-guided control prediction framework, the proposed \gls{dtcp} framework adaptively generates predicted commands based on real-time delay, thereby maintaining driving performance even with significant latency.

Extensive evaluations using the CARLA simulator demonstrate that the task-oriented co-design framework significantly improves driving performance under conditions of constrained bandwidth, noise interference, and varying communication delays. The proposed \gls{dtcp} consistently outperforms traditional methods across multiple scenarios. In particular, with an \gls{e2e} delay of 1 second (equivalent to 20 time slots in \gls{carla}), our framework achieves a driving score of 48.12, which is 31.59 points higher than when using BPG, while also reducing bandwidth usage by 99.19\%. Moreover, our analysis of compensating for perceived \gls{e2e} delay highlights the inherent unreliability of prediction under certain conditions, underscoring the need to balance predicted and unpredicted commands for optimal system performance. There are several promising directions for future research based on this work, such as extending the framework to more realistic wireless environments, including Urban Micro (UMi) and Urban Macro (UMa), and dynamically optimizing coding rates and modulation schemes based on channel conditions and SNR, leveraging 5G Modulation and Coding Scheme (MCS).

%% file: appendix.tex
\begin{appendices}
\section{Modeling Frequency-Selective Channel}
\label{apd_freq_selec}
We consider a multipath fading channel described by a discrete channel transfer function:
\begin{align}
    \check{\bm{z}}_{\text{time}} = \bm{h}_{\text{time}}* 
 \bm{z}_{\text{time}} + \bm{n}_{\text{time}},
 \label{eq_ofdm_time_domain}
\end{align}
where $*$ denotes the convolution operation. Here, $\check{\bm{z}}_{\text{time}}$ and $\bm{z}_{\text{time}}$ are the received and transmitted signals in the time domain, respectively, while $\bm{n}_{\text{time}}$ represents the additive Gaussian noise. The impulse response $\bm{h}_{\text{time}}=[h_{\text{time-}0}, \cdots, h_{\text{time-}(N_{\text{path}}-1)}]$ captures the multipath effect, where $h_{\text{time-}i}\sim \mathcal{CN}(0, \sigma_{i}^{2})$ for $i = 0, 1, \cdots, N_{\text{path}}-1$. We assume that path power decays exponentially as $\sigma_{i}^{2}=\alpha_i e^{-\frac{i}{\gamma}}$, with $\alpha_i$ ensuring power normalization $\sum_{i=0}^{N_{\text{path}}-1}\sigma_{i}^{2}=1$. Here, $\gamma$ is a delay spread constant.

To simplify, we assume synchronized transmission/reception without carrier frequency offset, and perfectly estimated channel state information by block-type pilot symbols. First, JSCC symbols $\bm{z} \in \mathbb{C}^{l_z}$ are padded with $N_\text{sub}-(l_z \bmod N_\text{sub})$ zeros and reshaped to $\bm{z}_r \in \mathbb{C}^{N_{\text{sym}}\times N_{\text{sub}}}$, where $N_\text{sym} = \lceil l_z / N_\text{sub} \rceil$ denotes the number of OFDM symbols and $N_\text{sub}$ represents the number of subcarriers per OFDM symbol. When $l_z / N_\text{sub}$ is not an integer, the subcarriers in the final OFDM symbol are not fully utilized for driving, but can be used for other tasks as needed.

Next, the Inverse Discrete Fourier Transform (IDFT) and cyclic prefix (CP) are applied, followed by transmission through the multipath channel as described in \cref{eq_ofdm_time_domain}. The receiver removes the CP and applies the Discrete Fourier Transform (DFT) to yield the received JSCC symbols $\check{\bm{z}}_r \in \mathbb{C}^{N_{\text{sym}}\times N_{\text{sub}}}$. Therefore, we have the following equation:
\begin{align} 
\check{\bm{z}}_{r}[j,k] = \bm{h}_{r}[j,k]\bm{z}_{r}[j,k] + \bm{n}_{r}[j,k], 
\label{eq_ofdm_freq_domain}
\end{align} 
where $k$ denotes the $k_{\text{th}}$ subcarrier, $j$ denotes the $j_\text{th}$ OFDM symbol, and $\bm{h}_{r}[j,k] = \bm{h}_{r}[j',k] , \forall j,j' \in \{1,\cdots,N_\text{sym}\}$ represents the subcarrier-specific channel response. Flatting each term and removing the dimensions of driving-irrelevant subcarriers lead to the simplified expression: 
\begin{align} 
\check{\bm{z}} = \bm{h} \cdot \bm{z} + \bm{n}, \notag 
\end{align} 
where $\check{\bm{z}} \in \mathbb{C}^{l_z}$, $\bm{h} \in \mathbb{C}^{l_z}$, and $\bm{n} \in \mathbb{C}^{l_z}$.

\section{Derivations of (\ref{eq_upper_bound_2nd})}
\label{apd_upper_bound_2nd}
The mutual information \(I(X; \hat{Z})\) can be written as:
\begin{align}
I(X ; \hat{Z}) & = \int p(\bm{x}, \hat{\bm{z}}) \log \frac{p(\bm{x}, \hat{\bm{z}})}{p(\bm{x}) p(\hat{\bm{z}})} \dif \bm{x} \dif \hat{\bm{z}} \notag\\
& = \int p(\bm{x}, \hat{\bm{z}}) \log \frac{p_{\phi}(\hat{\bm{z}}|\bm{x})}{p(\hat{\bm{z}})} \dif \bm{x} \dif \hat{\bm{z}}.
\end{align}
Since \(D_{\text{KL}}(p(\hat{\bm{z}})\parallel q(\hat{\bm{z}})) \geq 0 \), we have
\begin{align}
     \int p(\hat{\bm{z}})\log p(\hat{\bm{z}}) \dif \hat{\bm{z}} \geq \int p(\hat{\bm{z}})\log q(\hat{\bm{z}}) \dif \hat{\bm{z}}.
\end{align}
So that
\begin{align}
    I(X ; \hat{Z})  &\leq \int p(\bm{x}, \hat{\bm{z}}) \log \frac{p_{\phi}(\hat{\bm{z}}|\bm{x})}{q(\hat{\bm{z}})} \dif \bm{x} \dif \hat{\bm{z}} \notag\\
    & = \int p(\bm{a},\bm{x})p_{\phi}(\hat{\bm{z}}|\bm{x}) \log \frac{p_{\phi}(\hat{\bm{z}}|\bm{x})}{q(\hat{\bm{z}})} \dif \bm{a} \dif \bm{x} \dif \hat{\bm{z}} \notag\\
    & = \mathbb{E}_{\bm{a},\bm{x}}\left[D_{\text{KL}}(p_{\phi}(\hat{\bm{z}} | \bm{x}) \| q(\hat{\bm{z}})) \right].
\end{align}

\end{appendices}

%% file: reference.bib
@Article{Shao_2022_LTO,
  author   = {Shao, Jiawei and Mao, Yuyi and Zhang, Jun},
  journal  = {IEEE J. Sel. Areas Commun.},
  title    = {Learning Task-Oriented Communication for Edge Inference: An Information Bottleneck Approach},
  year     = {2022},
  number   = {1},
  pages    = {197-211},
  volume   = {40},
  doi      = {10.1109/JSAC.2021.3126087},
}

@InProceedings{Wu_2022_TgC,
  author    = {Wu, Penghao and Jia, Xiaosong and Chen, Li and Yan, Junchi and Li, Hongyang and Qiao, Yu},
  booktitle = {Proc. Adv. Neural Inf. Process. Syst.},
  title     = {Trajectory-guided Control Prediction for End-to-end Autonomous Driving: A Simple yet Strong Baseline},
  year      = {2022},
  editor    = {S. Koyejo and S. Mohamed and A. Agarwal and D. Belgrave and K. Cho and A. Oh},
  pages     = {6119--6132},
  publisher = {Curran Associates, Inc.},
  volume    = {35},
}

@InProceedings{Zhang_2021_EtE,
  author    = {Zhang, Zhejun and Liniger, Alexander and Dai, Dengxin and Yu, Fisher and Van Gool, Luc},
  booktitle = {Proc. Int. Conf. Comput. Vis.},
  title     = {End-to-End Urban Driving by Imitating a Reinforcement Learning Coach},
  year      = {2021},
  month     = oct,
  pages     = {15222-15232},
}

@Article{Bourtsoulatze_2019_DJS,
  author   = {Bourtsoulatze, Eirina and Burth Kurka, David and Gündüz, Deniz},
  journal  = {IEEE Trans. on Cogn. Commun. Netw.},
  title    = {Deep Joint Source-Channel Coding for Wireless Image Transmission},
  year     = {2019},
  number   = {3},
  pages    = {567-579},
  volume   = {5},
  doi      = {10.1109/TCCN.2019.2919300},
}

@Article{Kurka_2020_DfD,
  author   = {Kurka, David Burth and Gündüz, Deniz},
  journal  = {IEEE J. Sel. Areas Inf. Theory},
  title    = {{DeepJSCC}-f: Deep Joint Source-Channel Coding of Images With Feedback},
  year     = {2020},
  number   = {1},
  pages    = {178-193},
  volume   = {1},
  doi      = {10.1109/JSAIT.2020.2987203},
}

@Article{Shao_2023_TOC,
  author   = {Shao, Jiawei and Mao, Yuyi and Zhang, Jun},
  journal  = {IEEE Trans. Wireless Commun.},
  title    = {Task-Oriented Communication for Multidevice Cooperative Edge Inference},
  year     = {2023},
  number   = {1},
  pages    = {73-87},
  volume   = {22},
  doi      = {10.1109/TWC.2022.3191118},
}

@InProceedings{Li_2018_EIO,
  author    = {Li, En and Zhou, Zhi and Chen, Xu},
  booktitle = {Proc. Workshop Mobile Edge Commun.},
  title     = {Edge Intelligence: On-Demand Deep Learning Model Co-Inference with Device-Edge Synergy},
  year      = {2018},
  address   = {New York, NY, USA},
  pages     = {31–36},
  publisher = {Association for Computing Machinery},
  series    = {MECOMM'18},
  doi       = {10.1145/3229556.3229562},
  isbn      = {9781450359061},
  location  = {Budapest, Hungary},
  numpages  = {6},
  url       = {https://doi.org/10.1145/3229556.3229562},
}

@Article{Jankowski_2021_WIR,
  author   = {Jankowski, Mikolaj and Gündüz, Deniz and Mikolajczyk, Krystian},
  journal  = {IEEE J. Sel. Areas Commun.},
  title    = {Wireless Image Retrieval at the Edge},
  year     = {2021},
  number   = {1},
  pages    = {89-100},
  volume   = {39},
  doi      = {10.1109/JSAC.2020.3036955},
}

@InProceedings{Tishby_1999_TIB,
  author    = {Tishby, Naftali and Pereira, Fernando C and Bialek, William},
  booktitle = {Proc. Annu. Allerton Conf. Commun. Control Comput.},
  title     = {The Information Bottleneck Method},
  year      = {1999},
  pages     = {368–377},
}

@InProceedings{Kurka_2019_SRo,
  author    = {Kurka, David Burth and Gündüz, Deniz},
  booktitle = {Proc. IEEE Int. Workshop Signal Process. Adv. Wireless Commun.},
  title     = {Successive Refinement of Images with Deep Joint Source-Channel Coding},
  year      = {2019},
  pages     = {1-5},
  doi       = {10.1109/SPAWC.2019.8815416},
}

@Article{Wallace_1992_TJs,
  author   = {Wallace, G.K.},
  journal  = {IEEE Trans. Consum. Electron.},
  title    = {The {JPEG} still picture compression standard},
  year     = {1992},
  number   = {1},
  pages    = {18-34},
  volume   = {38},
  doi      = {10.1109/30.125072},
}

@Article{Taubman_2002_JIc,
  author    = {Taubman, David S and Marcellin, Michael W and Rabbani, Majid},
  journal   = {J. Electron. Imag.},
  title     = {{JPEG2000}: Image compression fundamentals, standards and practice},
  year      = {2002},
  number    = {2},
  pages     = {286--287},
  volume    = {11},
  publisher = {Society of Photo-Optical Instrumentation Engineers},
}

@Article{Yang_2022_OGD,
  author   = {Yang, Mingyu and Bian, Chenghong and Kim, Hun-Seok},
  journal  = {IEEE Trans. Cogn. Commun. Netw.},
  title    = {{OFDM}-Guided Deep Joint Source Channel Coding for Wireless Multipath Fading Channels},
  year     = {2022},
  number   = {2},
  pages    = {584-599},
  volume   = {8},
  doi      = {10.1109/TCCN.2022.3151935},
}

@Article{Gallager_1962_Ldp,
  author   = {Gallager, R.},
  journal  = {IRE Trans. Inf. Theory},
  title    = {Low-density parity-check codes},
  year     = {1962},
  number   = {1},
  pages    = {21-28},
  volume   = {8},
  doi      = {10.1109/TIT.1962.1057683},
}

@InProceedings{Alemi_2017_DVI,
  author    = {Alexander A. Alemi and Ian Fischer and Joshua V. Dillon and Kevin Murphy},
  booktitle = {Proc. Int. Conf. Learn. Represent.},
  title     = {Deep Variational Information Bottleneck},
  year      = {2017},
  url       = {https://openreview.net/forum?id=HyxQzBceg},
}

@InProceedings{Shao_2020_BAE,
  author    = {Shao, Jiawei and Zhang, Jun},
  booktitle = {Proc. IEEE Int. Conf. Commun. Workshops},
  title     = {{BottleNet}++: An End-to-End Approach for Feature Compression in Device-Edge Co-Inference Systems},
  year      = {2020},
  pages     = {1-6},
  doi       = {10.1109/ICCWorkshops49005.2020.9145068},
}

@Article{Shi_2020_CEE,
  author   = {Shi, Yuanming and Yang, Kai and Jiang, Tao and Zhang, Jun and Letaief, Khaled B.},
  journal  = {IEEE Commun. Surv. Tut.},
  title    = {Communication-Efficient Edge {AI}: Algorithms and Systems},
  year     = {2020},
  number   = {4},
  pages    = {2167-2191},
  volume   = {22},
  doi      = {10.1109/COMST.2020.3007787},
}

@Article{Huang_2020_DCR,
  author   = {Huang, Xiufeng and Zhou, Sheng},
  journal  = {IEEE Internet Things J.},
  title    = {Dynamic Compression Ratio Selection for Edge Inference Systems With Hard Deadlines},
  year     = {2020},
  number   = {9},
  pages    = {8800-8810},
  volume   = {7},
  doi      = {10.1109/JIOT.2020.2997128},
}

@InProceedings{Shi_2019_IDE,
  author    = {Shi, Wenqi and Hou, Yunzhong and Zhou, Sheng and Niu, Zhisheng and Zhang, Yang and Geng, Lu},
  booktitle = {Proc. IEEE Conf. Comput. Commun. Workshops},
  title     = {Improving Device-Edge Cooperative Inference of Deep Learning via 2-Step Pruning},
  year      = {2019},
  pages     = {1-6},
  doi       = {10.1109/INFOCOMWKSHPS47286.2019.9093772},
}

@Article{Li_2020_EAO,
  author   = {Li, En and Zeng, Liekang and Zhou, Zhi and Chen, Xu},
  journal  = {IEEE Trans. Wireless Commun.},
  title    = {Edge {AI}: On-Demand Accelerating Deep Neural Network Inference via Edge Computing},
  year     = {2020},
  number   = {1},
  pages    = {447-457},
  volume   = {19},
  doi      = {10.1109/TWC.2019.2946140},
}

@Article{Shao_2020_CCT,
  author   = {Shao, Jiawei and Zhang, Jun},
  journal  = {IEEE Commun. Mag.},
  title    = {Communication-Computation Trade-off in Resource-Constrained Edge Inference},
  year     = {2020},
  number   = {12},
  pages    = {20-26},
  volume   = {58},
  doi      = {10.1109/MCOM.001.2000373},
}

@InProceedings{Jankowski_2020_JDE,
  author    = {Jankowski, Mikolaj and Gündüz, Deniz and Mikolajczyk, Krystian},
  booktitle = {Proc. IEEE Int. Workshop Signal Process. Adv. Wireless Commun.},
  title     = {Joint Device-Edge Inference over Wireless Links with Pruning},
  year      = {2020},
  pages     = {1-5},
  doi       = {10.1109/SPAWC48557.2020.9154306},
}

@InProceedings{Dubois_2021_LCf,
  author    = {Dubois, Yann and Bloem-Reddy, Benjamin and Ullrich, Karen and Maddison, Chris J},
  booktitle = {Proc. Adv. Neural Inf. Process. Syst.},
  title     = {Lossy Compression for Lossless Prediction},
  year      = {2021},
  editor    = {M. Ranzato and A. Beygelzimer and Y. Dauphin and P.S. Liang and J. Wortman Vaughan},
  pages     = {14014--14028},
  publisher = {Curran Associates, Inc.},
  volume    = {34},
  url       = {https://proceedings.neurips.cc/paper_files/paper/2021/file/7535bbb91c8fde347ad861f293126633-Paper.pdf},
}

@InProceedings{Shao_2021_BGA,
  author    = {Shao, Jiawei and Zhang, Haowei and Mao, Yuyi and Zhang, Jun},
  booktitle = {Proc. IEEE Int. Conf. Acoust., Speech, Signal Process.},
  title     = {{Branchy-GNN}: A Device-Edge Co-Inference Framework for Efficient Point Cloud Processing},
  year      = {2021},
  pages     = {8488-8492},
  doi       = {10.1109/ICASSP39728.2021.9414831},
}

@InProceedings{Tishby_2015_Dla,
  author    = {Tishby, Naftali and Zaslavsky, Noga},
  booktitle = {Proc. IEEE Inf. Theory Workshop},
  title     = {Deep learning and the information bottleneck principle},
  year      = {2015},
  pages     = {1-5},
  doi       = {10.1109/ITW.2015.7133169},
}

@Article{Saidutta_2021_JSC,
  author   = {Saidutta, Yashas Malur and Abdi, Afshin and Fekri, Faramarz},
  journal  = {IEEE J. Sel. Areas Commun.},
  title    = {Joint Source-Channel Coding Over Additive Noise Analog Channels Using Mixture of Variational Autoencoders},
  year     = {2021},
  number   = {7},
  pages    = {2000-2013},
  volume   = {39},
  doi      = {10.1109/JSAC.2021.3078489},
}

@Article{Chaccour_2024_LDM,
  author   = {Chaccour, Christina and Saad, Walid and Debbah, Mérouane and Han, Zhu and Poor, H. Vincent},
  journal  = {IEEE Commun. Surv. Tut.},
  title    = {Less Data, More Knowledge: Building Next Generation Semantic Communication Networks},
  year     = {2024},
  pages    = {1-1},
  doi      = {10.1109/COMST.2024.3412852},
}

@Article{Yang_2023_SCf,
  author   = {Yang, Wanting and Du, Hongyang and Liew, Zi Qin and Lim, Wei Yang Bryan and Xiong, Zehui and Niyato, Dusit and Chi, Xuefen and Shen, Xuemin and Miao, Chunyan},
  journal  = {IEEE Commun. Surv. Tut.},
  title    = {Semantic Communications for Future Internet: Fundamentals, Applications, and Challenges},
  year     = {2023},
  number   = {1},
  pages    = {213-250},
  volume   = {25},
  doi      = {10.1109/COMST.2022.3223224},
}

@Online{BPG,
  author = {Bellard, Fabrice},
  title  = {{BPG} Image Format},
  url    = {https://bellard.org/bpg/},
  year   = {2014},
}

@InProceedings{Pandey_2023_GOC,
  author    = {Pandey, Shashi Raj and Bui, Van Phuc and Popovski, Petar},
  booktitle = {Proc. IEEE Int. Symp. Pers. Indoor Mob. Radio Commun.},
  title     = {Goal-Oriented Communications in Federated Learning via Feedback on Risk-Averse Participation},
  year      = {2023},
  pages     = {1-6},
  doi       = {10.1109/PIMRC56721.2023.10293926},
}

@Article{Kang_2023_PSi,
  author   = {Kang, Jiawen and Du, Hongyang and Li, Zonghang and Xiong, Zehui and Ma, Shiyao and Niyato, Dusit and Li, Yuan},
  journal  = {IEEE J. Sel. Areas Commun.},
  title    = {Personalized Saliency in Task-Oriented Semantic Communications: Image Transmission and Performance Analysis},
  year     = {2023},
  number   = {1},
  pages    = {186-201},
  volume   = {41},
  doi      = {10.1109/JSAC.2022.3221990},
}

@Article{Arimoto_1972_Aaf,
  author  = {Arimoto, S.},
  journal = {IEEE Trans. Inf. Theory},
  title   = {An algorithm for computing the capacity of arbitrary discrete memoryless channels},
  year    = {1972},
  number  = {1},
  pages   = {14-20},
  volume  = {18},
  doi     = {10.1109/TIT.1972.1054753},
}

@Article{Blahut_1972_Coc,
  author  = {Blahut, R.},
  journal = {IEEE Trans. Inf. Theory},
  title   = {Computation of channel capacity and rate-distortion functions},
  year    = {1972},
  number  = {4},
  pages   = {460-473},
  volume  = {18},
  doi     = {10.1109/TIT.1972.1054855},
}

@InProceedings{Strinati_2024_GOa,
  author        = {Strinati, Emilio Calvanese and Di Lorenzo, Paolo and Sciancalepore, Vincenzo and Aijaz, Adnan and Kountouris, Marios and Gündüz, Deniz and Popovski, Petar and Sana, Mohamed and Stavrou, Photios A. and Soret, Beatriz and Cordeschi, Nicola and Scardapane, Simone and Merluzzi, Mattia and Zanzi, Lanfranco and Renato, Mauro Boldi and Quek, Tony and Pietro, Nicola Di and Forceville, Olivier and Costanzo, Francesca and Li, Peizheng},
  booktitle     = {Proc. Joint Eur. Conf. Netw. Commun. 6G Summit},
  title         = {Goal-Oriented and Semantic Communication in {6G} {AI}-Native Networks: The {6G-GOALS} Approach},
  year          = {2024},
  pages         = {1-6},
  comment-eidos = {2024 Joint European Conference on Networks and Communications & 6G Summit (EuCNC/6G Summit)},
  doi           = {10.1109/EuCNC/6GSummit60053.2024.10597087},
}

@Article{Kingma_2013_Aev,
  author  = {Kingma, Diederik P and Welling, Max},
  journal = {arXiv preprint arXiv:1312.6114},
  title   = {Auto-encoding variational bayes},
  year    = {2013},
}

@InProceedings{Cho_2014_LPR,
  author    = {Cho, Kyunghyun and van Merri{\"e}nboer, Bart and Gulcehre, Caglar and Bahdanau, Dzmitry and Bougares, Fethi and Schwenk, Holger and Bengio, Yoshua},
  booktitle = {Proc. Conf. Empirical Methods Nat. Lang. Process.},
  title     = {Learning Phrase Representations using {RNN} Encoder{--}Decoder for Statistical Machine Translation},
  year      = {2014},
  address   = {Doha, Qatar},
  editor    = {Moschitti, Alessandro and Pang, Bo and Daelemans, Walter},
  month     = oct,
  pages     = {1724--1734},
  publisher = {Association for Computational Linguistics},
  doi       = {10.3115/v1/D14-1179},
  url       = {https://aclanthology.org/D14-1179/},
}

@Article{Hou_2020_MPa,
  author   = {Hou, Xueshi and Dey, Sujit},
  journal  = {IEEE Open J. Commun. Soc.},
  title    = {Motion Prediction and Pre-Rendering at the Edge to Enable Ultra-Low Latency Mobile {6DoF} Experiences},
  year     = {2020},
  pages    = {1674-1690},
  volume   = {1},
  doi      = {10.1109/OJCOMS.2020.3032608},
}

@InProceedings{Richter_2019_ARP,
  author    = {Richter, Florian and Zhang, Yifei and Zhi, Yuheng and Orosco, Ryan K. and Yip, Michael C.},
  booktitle = {Proc. Int. Conf. Robot. Autom.},
  title     = {Augmented Reality Predictive Displays to Help Mitigate the Effects of Delayed Telesurgery},
  year      = {2019},
  pages     = {444-450},
  doi       = {10.1109/ICRA.2019.8794051},
}

@InProceedings{Tong_2018_MWR,
  author    = {Tong, Xin and Zhao, Guodong and Imran, Muhammad Ali and Pang, Zhibo and Chen, Zhi},
  booktitle = {Proc. IEEE Int. Conf. Commun. Workshops},
  title     = {Minimizing Wireless Resource Consumption for Packetized Predictive Control in Real-Time Cyber Physical Systems},
  year      = {2018},
  pages     = {1-6},
  doi       = {10.1109/ICCW.2018.8403546},
}

@Article{Simsek_2016_5ET,
  author   = {Simsek, Meryem and Aijaz, Adnan and Dohler, Mischa and Sachs, Joachim and Fettweis, Gerhard},
  journal  = {IEEE J. Sel. Areas Commun.},
  title    = {{5G}-Enabled Tactile Internet},
  year     = {2016},
  number   = {3},
  pages    = {460-473},
  volume   = {34},
  doi      = {10.1109/JSAC.2016.2525398},
}

@Article{Hou_2020_PaC,
  author   = {Hou, Zhanwei and She, Changyang and Li, Yonghui and Zhuo, Li and Vucetic, Branka},
  journal  = {IEEE Trans. Wireless Commun.},
  title    = {Prediction and Communication Co-Design for Ultra-Reliable and Low-Latency Communications},
  year     = {2020},
  number   = {2},
  pages    = {1196-1209},
  volume   = {19},
  doi      = {10.1109/TWC.2019.2951660},
}

@Article{Meng_2023_SCa,
  author   = {Meng, Zhen and She, Changyang and Zhao, Guodong and De Martini, Daniele},
  journal  = {IEEE J. Sel. Areas Commun.},
  title    = {Sampling, Communication, and Prediction Co-Design for Synchronizing the Real-World Device and Digital Model in Metaverse},
  year     = {2023},
  number   = {1},
  pages    = {288-300},
  volume   = {41},
  doi      = {10.1109/JSAC.2022.3221993},
}

@Article{She_2019_CLD,
  author   = {She, Changyang and Duan, Yifan and Zhao, Guodong and Quek, Tony Q. S. and Li, Yonghui and Vucetic, Branka},
  journal  = {IEEE Internet Things J.},
  title    = {Cross-Layer Design for Mission-Critical {IoT} in Mobile Edge Computing Systems},
  year     = {2019},
  number   = {6},
  pages    = {9360-9374},
  volume   = {6},
  doi      = {10.1109/JIOT.2019.2930983},
}

@Article{She_2021_ATo,
  author   = {She, Changyang and Sun, Chengjian and Gu, Zhouyou and Li, Yonghui and Yang, Chenyang and Poor, H. Vincent and Vucetic, Branka},
  journal  = {Proc. IEEE},
  title    = {A Tutorial on Ultrareliable and Low-Latency Communications in {6G}: Integrating Domain Knowledge Into Deep Learning},
  year     = {2021},
  number   = {3},
  pages    = {204-246},
  volume   = {109},
  doi      = {10.1109/JPROC.2021.3053601},
}

@Article{Meng_2024_TOC,
  author   = {Meng, Zhen and Chen, Kan and Diao, Yufeng and She, Changyang and Zhao, Guodong and Imran, Muhammad Ali and Vucetic, Branka},
  journal  = {IEEE J. Sel. Areas Commun.},
  title    = {Task-Oriented Cross-System Design for Timely and Accurate Modeling in the Metaverse},
  year     = {2024},
  number   = {3},
  pages    = {752-766},
  volume   = {42},
  doi      = {10.1109/JSAC.2023.3345398},
}

@Article{Khan_2022_Uae,
  author   = {Khan, Benish Sharfeen and Jangsher, Sobia and Ahmed, Ashfaq and Al-Dweik, Arafat},
  journal  = {IEEE Open J. Commun. Soc.},
  title    = {{URLLC} and {eMBB} in {5G} Industrial {IoT}: A Survey},
  year     = {2022},
  pages    = {1134-1163},
  volume   = {3},
  doi      = {10.1109/OJCOMS.2022.3189013},
}

@InProceedings{Diao_2024_TOS,
  author    = {Diao, Yufeng and Meng, Zhen and Xu, Xiangmin and She, Changyang and Zhao, Philip G.},
  booktitle = {Proc. IEEE Conf. Comput. Commun. Workshops},
  title     = {Task-Oriented Source-Channel Coding Enabled Autonomous Driving Based on Edge Computing},
  year      = {2024},
  pages     = {1-6},
  doi       = {10.1109/INFOCOMWKSHPS61880.2024.10620735},
}

@InProceedings{Liao_2024_AAG,
  author    = {Liao, Qi and Tung, Tze-Yang},
  booktitle = {Proc. IEEE Conf. Comput. Commun.},
  title     = {{AdaSem}: Adaptive Goal-Oriented Semantic Communications for End-to-End Camera Relocalization},
  year      = {2024},
  pages     = {1111-1120},
  doi       = {10.1109/INFOCOM52122.2024.10621397},
}

@Article{Chen_2024_SMW,
  author   = {Chen, Long and Li, Yuchen and Silamu, Wushour and Li, Qingquan and Ge, Shirong and Wang, Fei-Yue},
  journal  = {IEEE Transactions on Intelligent Vehicles},
  title    = {Smart Mining With Autonomous Driving in Industry 5.0: Architectures, Platforms, Operating Systems, Foundation Models, and Applications},
  year     = {2024},
  number   = {3},
  pages    = {4383-4393},
  volume   = {9},
  doi      = {10.1109/TIV.2024.3365997},
}

@Article{3GPP,
  journal = {\rm{document 3GPP, TSG RAN TR38.913 R14}},
  title   = {\emph{Study on scenarios and requirements for next generation access technologies}},
  year    = {2017},
  month   = Jun,
}

@InProceedings{Diao_2024_TTG,
  author    = {Diao, Yufeng and Zhang, Yichi and Zhao, Philip G. and De Martini, Daniele},
  booktitle = {Proc. IEEE Conf. Comput. Commun. Workshops},
  title     = {{TAGIC}: Task-Guided Image Communication Framework for Seamless Teleoperation},
  year      = {2024},
  pages     = {1-2},
  doi       = {10.1109/INFOCOMWKSHPS61880.2024.10620864},
}

@InProceedings{Molchanov_2017_VDS,
  author    = {Dmitry Molchanov and Arsenii Ashukha and Dmitry Vetrov},
  booktitle = {Proc. 34th Int. Conf. Mach. Learn.},
  title     = {Variational Dropout Sparsifies Deep Neural Networks},
  year      = {2017},
  editor    = {Precup, Doina and Teh, Yee Whye},
  pages     = {2498--2507},
  publisher = {PMLR},
  series    = {Proceedings of Machine Learning Research},
  volume    = {70},
  url       = {https://proceedings.mlr.press/v70/molchanov17a.html},
}

@TechReport{GPP_2020_5NP,
  author      = {{3rd Generation Partnership Project (3GPP)}},
  institution = {3GPP},
  title       = {{5G; NR; Physical Channels and Modulation (3GPP TS 38.211 version 16.2.0 Release 16)}},
  year        = {2020},
  note        = {[Online]. Available: https://www.3gpp.org},
}

@Article{Guenduez_2023_BTB,
  author   = {Gündüz, Deniz and Qin, Zhijin and Aguerri, Inaki Estella and Dhillon, Harpreet S. and Yang, Zhaohui and Yener, Aylin and Wong, Kai Kit and Chae, Chan-Byoung},
  journal  = {IEEE J. Sel. Areas Commun.},
  title    = {Beyond Transmitting Bits: Context, Semantics, and Task-Oriented Communications},
  year     = {2023},
  number   = {1},
  pages    = {5-41},
  volume   = {41},
  doi      = {10.1109/JSAC.2022.3223408},
}
